\documentclass[prb,twocolumn,showpacs,amsmath,amssymb,superscriptaddress]{revtex4-1}
\usepackage{graphicx}
\usepackage{color}
\usepackage{braket}
\usepackage{amsmath}
\usepackage{amsfonts}
\usepackage{MnSymbol}

\usepackage{amsmath,amssymb,mathrsfs,ulem}
\usepackage[colorlinks = true,linkcolor = blue,urlcolor  = blue,     citecolor = blue,anchorcolor = blue]{hyperref}
\usepackage{bbm}
\usepackage[dvipsnames]{xcolor}
\usepackage{hyperref}

\definecolor{amber}{rgb}{1.0, 0.75, 0.0}

\date{\today}                  

\begin{document}

\title{Universal properties of boundary and interface charges in multichannel one-dimensional continuum models }

\author{Kiryl Piasotski}
\affiliation{Institut f\"ur Theorie der Statistischen Physik, RWTH Aachen, 
52056 Aachen, Germany and JARA - Fundamentals of Future Information Technology}

\author{Niclas M\"uller}
\affiliation{Institut f\"ur Theorie der Statistischen Physik, RWTH Aachen, 
52056 Aachen, Germany and JARA - Fundamentals of Future Information Technology}

\author{Dante M. Kennes}
\affiliation{Institut f\"ur Theorie der Statistischen Physik, RWTH Aachen, 
52056 Aachen, Germany and JARA - Fundamentals of Future Information Technology}
\affiliation{Max Planck Institute for the Structure and Dynamics of Matter, Center for Free Electron Laser Science, 22761 Hamburg, Germany}

\author{Herbert Schoeller}
\affiliation{Institut f\"ur Theorie der Statistischen Physik, RWTH Aachen, 
52056 Aachen, Germany and JARA - Fundamentals of Future Information Technology}

\author{Mikhail Pletyukhov}
\email[Email: ]{pletmikh@physik.rwth-aachen.de}
\affiliation{Institut f\"ur Theorie der Statistischen Physik, RWTH Aachen, 
52056 Aachen, Germany and JARA - Fundamentals of Future Information Technology}

\begin{abstract}
We generalize our recent results for the boundary and interface charges in one-dimensional single-channel continuum [Phys. Rev. B {\bf 104}, 155409 (2021)] and multichannel tight-binding [Phys. Rev. B {\bf 104}, 125447 (2021)] models to the realm of the multichannel continuum systems. Using the technique of boundary Green's functions, we give a rigorous proof that the change in boundary charge upon the shift of the system towards the boundary by the distance $x_{\varphi}\in[0, L]$ is given by a perfectly linear function of $x_{\varphi}$ plus an integer-valued topological invariant $I$ -- the so called boundary invariant. We provide two equivalent representations for $I(x_{\varphi})$: the winding number representation and the bound state representation. The winding number representation allows one to write $I$ as a winding index of a particular functional of bulk Green's function. The corresponding integration contour is chosen in  the complex frequency plane to encircle the occupied part of the spectrum residing on the real axis. In turn, in the bound state representation,  $I$ is expressed through the sum of the winding number of the boundary Green's function and the number of bound states supported by the cavity of size $x_{\varphi}$ below the chemical potential. We observe that during a single cycle in the variation of $x_{\varphi}$, the boundary invariant exhibits exactly $\nu$ downward jumps, each by a unit of electron charge, whenever $\nu$ energy bands are completely filled leading to the value $I(L)=-\nu$. Additionally, for translationally invariant models interrupted by a localized impurity we derive the winding number expression for the excess charge accumulated on the said impurity. We observe that the charge accumulated on a single repulsive impurity is restricted to the values $-N_{c}, ..., 0$, where $N_{c}$ is the number of channels (spin or orbital components) in the system. For systems with weak potential amplitudes, we additionally develop Green's function-based low-energy theory, allowing one to analytically access the physics of multichannel continuum systems in the low-energy approximation.

\end{abstract}


\maketitle

\section{Introduction}
Over the last few decades, the notion of topology has swiftly moved to the frontiers of contemporary condensed matter research [\onlinecite{Laughlin_1981, Thouless_1982, Thouless_1983, Kohmoto_1985, Haldane_1988, Hatsugai_1993, Kane_2005, Kane2_2005, Bernevig_2006, Bernevig2_2006, Koenig_2007}]. Possibly, the most prominent example of the interplay between topology and physics are the so-called topological insulators (TI) - insulators with gapped bulk and topologically protected gapless edge states [\onlinecite{He_2019}]. Specifically, the very presence of these edge modes has motivated an increasing interest in the field of TIs due to their promising applications in quantum technology [\onlinecite{Miyake_2010, Pesin_2012, Stern_2013, Mong_2014, Bomantara_2018}]. In a TI, these edge states are said to be symmetry-protected when the said insulator falls into one of the ten discrete symmetry classes [\onlinecite{Kitaev_2009}] according to its transformation properties under time-reversal, particle-hole, and chiral symmetries [\onlinecite{Altland_1997}]. To each of the respective symmetry classes, there exists an associated topological invariant (such as a Chern number for example), allowing one to distinguish between trivial (edge states are absent) and topological (edge states are present) phases [\onlinecite{Hasan_2010}, \onlinecite{Bernevig_2013}]. This picture naturally suggests two peculiar physical questions. Firstly, whether or not topological invariants may be directly related to a certain physically observable quantity and, secondly, whether it is possible to establish a link between topology and solid-state systems going beyond the constraints of symmetries? 
\par 
The first attempt to answer these questions appeared with the advent of the modern theory of polarization (MTP) [\onlinecite{King_1993}]. In particular, the so-called surface charge theorem was put forward relating the boundary charge $Q_{B}$ (the excess charge accumulated on the boundary of an insulator) to the Zak-Berry phase of the bulk Bloch states [\onlinecite{Vanderbilt_1993, Vanderbilt_2012, Vanderbilt_2018}]. Despite the connection between the boundary charge and the Zak-Berry phase not being subject to any symmetry restrictions, this connection lacks uniqueness. Indeed, this relation holds only up to an integer since, upon a gauge transformation, the Zak-Berry phase changes by a winding number of this transformation [\onlinecite{Vanderbilt_2018}]. This observation recently served as one of the principal motivations to study the properties of the boundary charge in one-dimensional insulators in more detail [\onlinecite{Park_2016}, \onlinecite{Thakurathi_2018}].  
\par
In a series of our recent works [\onlinecite{Pletyukhov_2020, Pletyukhov_2020_R, Pletyukhov_2020_prr, Weber_2021, Piasotski_2021, Muller_2021, Miles_21}], a number of the universal features of boundary $Q_{B}$ and interface $Q_{I}$ (the excess charge accumulated on a localized impurity or at interface between two insulators) charges was established. In Ref. [\onlinecite{Pletyukhov_2020}, \onlinecite{Pletyukhov_2020_R}] topological indices related to the boundary charge in a broad class of generalized Aubry-Andr\'e-Harper models (models defined by the set of periodic hopping $t_{m}=t_{m+Z}$ and potential $v_{m}=v_{m+Z}$ amplitudes [\onlinecite{Harper_55}, \onlinecite{Aubry_1980}], with $Z$ being the number of sites in a unit cell) were introduced. Specifically, a particular gauge has been identified, in which the relation between the Zak-Berry phase and $Q_{B}$ becomes exact, fixing the unknown integer of the surface charge theorem of MTP. Furthermore, an integer-valued symmetry-independent topological invariant $I\in\{-1, 0\}$ (so-called boundary invariant) associated with the change in the boundary charge upon the shift of the lattice by a single site towards the boundary ($t_{m}\rightarrow t_{m+1}, \ v_{m}\rightarrow v_{m+1}$) was introduced. In addition, it was demonstrated that upon a continuous shift of the lattice towards the boundary by the change in the phase of the modulation of the model parameters, the total boundary charge consists of (up to a finite number of discontinuous jumps) a universal linear function of the modulation phase and a non-universal $\frac{2\pi}{Z}$ periodic function vanishing in the $Z\gg1$ limit. 
\par 
In Ref. [\onlinecite{Pletyukhov_2020_prr}], the universal properties of $Q_{B}$ and $Q_{I}$ were put into a general framework based on the nearsightedness principle of W. Kohn [\onlinecite{Kohn_1996}, \onlinecite{Prodan_2005}] stating that localized perturbations in insulators can only lead to local charge redistributions.  In particular, based on this principle, two quantized invariants were established for generic one-dimensional tight-binding models (including the multichannel models -- models with multiple orbitals per site). The first invariant, in analogy with the single-channel case [\onlinecite{Pletyukhov_2020}, \onlinecite{Pletyukhov_2020_R}], underpins the universal behavior of the boundary charge upon discrete lattice translations. The second one is related to the local inversion of the lattice and characterizes the sum of the two boundary charges left and right to the septum between two insulators. These developments form a basis for the theory of the interface charges on localized impurities and domain walls and provide a generalization of the Goldstone-Wilczek formula [\onlinecite{Jackiw_1976, Su_1980, Rice_1982, Heeger_1988, Goldstone_1981}].
\par
Quite recently, general conclusions of Ref.~[\onlinecite{Pletyukhov_2020_prr}] were given a rigorous mathematical proof [\onlinecite{Muller_2021}]. In particular, for arbitrary translationally invariant tight-binding models with $N_{c}$ channels (orbitals) per site, the boundary invariant was shown to be given by the winding number of a given functional of bulk Green's functions, providing an extended form of the bulk-boundary correspondence. As opposed to the single-channel case, $I$ is no longer restricted to the values $-1, 0$ but rather takes on values in the range $-N_{c}, ..., 0$, naturally generalizing the $N_{c}=1$ case. Likewise, the quantization of the interface charge on isolated lattice defects was demonstrated by establishing the winding number expressions for the corresponding interface charges.
\par 
In another recent work, the continuum limit of the Aubry-Andr\'e-Harper models was analyzed [\onlinecite{Miles_21}]. Such a continuum limit may be achieved by taking the simultaneous limits $Z\rightarrow\infty$ and $a\rightarrow0$ (with $a$ being the lattice spacing), while keeping the size of the unit cell finite $Za=L$, in this case, one arrives at the conventional Schr\"odinger equation for a particle of mass $m=\frac{\hbar^{2}}{2ta^{2}}$ moving in a periodic scalar potential $V(x)=V(x+L)$. In analogy with the tight-binding models the change in the boundary charge upon the system's shift was studied, albeit now, instead of a shift of the system by a single lattice site, the shift of the lattice as a whole towards the boundary by a continuous coordinate $x_{\varphi}\in[0, L]$ was considered. It was revealed that the change in the total boundary charge, up to a number of jumps by minus one electron charge at the points where an edge state crosses the chemical potential, is given by a perfectly linear function of $x_{\varphi}$ with the slope given by $\nu/L$. The study thus completely confirms the observation of [\onlinecite{Pletyukhov_2020}, \onlinecite{Pletyukhov_2020_R}] that the non-universal contribution to the boundary charge vanishes exactly in the $Z\rightarrow\infty$ limit. Further, it was revealed that in the case of $\nu$ occupied bands, the change in the boundary charge exhibits exactly $\nu$ downward jumps by a unit of electron charge, implying that in the continuum models, the $\nu$th band gap hosts exactly $\nu$ (either left- or right-sided) edge states. In addition, following [\onlinecite{Muller_2021}], integer charge quantization on localized impurities was demonstrated. 
\par
In this paper we study continuum models of one-dimensional multichannel insulators beyond any symmetry constraints. The models we focus on are defined by all one-dimensional Schr\"odinger operators with periodic non-Abelian scalar $V(x)=V(x+L)$ and vector $A(x)=A(x+L)$ potentials (here $V$ and $A$ are assumed to be $N_{c}\times N_{c}$ hermitian matrices). This class of models describes a wide variety of interesting physical systems. For example, non-Abelian gauge fields may naturally arise from relativistic corrections such as Zeeman and spin-orbit terms in quantum nanowires [\onlinecite{Weperen_2015}, \onlinecite{Kloeffel_2011}]. These systems usually possess topologically non-trivial band structures [\onlinecite{Rainis_2013, Klinovaja_2012, Maier_2014, Klinovaja_2012_prl}], and have a great potential for spintronics applications [\onlinecite{Liang_2012, Nautiyal_2004, Iorio_2018}]. Famously, non-Abelian fields arise in the adiabatic dynamics of degenerate quantum systems (generalizing the idea of the Berry phase) as was shown by Wilczek and Zee [\onlinecite{Wilczek_1984}]. Another prominent example is given by cold-atom systems, where the artificial (or effective) non-Abelian gauge fields arise from the motion of multilevel atoms in spatially varying laser fields which couple the states in the degenerate subspaces between one another [\onlinecite{Juzeliunas_2008, Ruseckas_2005, Burrello_2010, Satija_2006, Pietila_2009}]; see for example [\onlinecite{Lin_2009}] for an experimental realization. Other noteworthy applications include the emergent non-Abelian gauge fields in twisted graphene bilayers [\onlinecite{Jose_2012}, \onlinecite{Yin_2015}], trapped fermion systems [\onlinecite{Ghosh_2011}], and semiconductor photonic cavities [\onlinecite{Tercas_2014}].
\par
Since the bands in multichannel systems are typically composite (see for example [\onlinecite{Marzari_1997}] for definition), the problem of exact diagonalization of the semi-infinite system's Hamiltonian in the spirit of [\onlinecite{Pletyukhov_2020}, \onlinecite{Miles_21}] becomes rather challenging. In order to avoid these complications, following [\onlinecite{Muller_2021}], we utilize the method of boundary Green's functions (BGF) [\onlinecite{Komnik_etal_2017}]. The BGF technique is common in various transport calculations including, for example, electronic transport in metals [\onlinecite{Zazunov_etal_2017}, \onlinecite{Zazunov_etal_2018}] and thermal transport in spin heterostructures [\onlinecite{Arrachea_etal_2009}]. Lately this method was employed in the realm of topological insulators [\onlinecite{Peng_etal_2017}] and superconductors [\onlinecite{Zazunov_etal_2016}, \onlinecite{Alvarado_etal_2020}].
\par
This work is dedicated to the study of both boundary and interface charges. As it was mentioned above, we employ the BGF method to study boundary and interface charge-based topological invariants. The boundary invariant is defined in the manner analogous to the single-channel continuum models [\onlinecite{Miles_21}], with a continuous shift parameter $x_{\varphi}\in[0, L]$. We give a rigorous proof that the boundary invariant associated with the corresponding shift is a quantized integer quantity and provide two different representations for the invariant. The first representation is based on the winding number expression. The quantity whose winding determines $I$ is shown to be a functional of the equal-argument spatial bulk Green's function and its derivatives with respect to both spatial arguments. Since the change in the boundary charge jumps by an integer only when an edge state crosses the chemical potential from below, this invariant naturally provides a form of the bulk-boundary correspondence in multichannel continuum systems. That is the spectral flow of boundary (or edge) modes is completely characterized by the bulk quantities. The second representation, on the other hand, utilizes the boundary Green's functions solely. In this representation, the invariant is expressed as a sum of the winding number of BGF evaluated at $x=x'=x_{\varphi}$ and the number of bound states supported by the cavity (or box) formed by two hard walls at $x=0$ and $x_{\varphi}$. What concerns the interface charge, we study the translationally invariant system interrupted by a single isolated ($\delta$-like) repulsive impurity and show that the excess charge accumulated on the impurity is an integer multiple of the electron charge and is given by a winding number expression completely analogous to the one found in the tight-biding realization of the model [\onlinecite{Muller_2021}]. 
\par
We substantiate our analytical findings by numerical analysis of both boundary and interface invariants. In our analysis we decompose $S(x)=\bar{S}(x)+\delta S(x)$, where $\bar{S}(x)$ and $\delta{S}(x)$ are the Abelian and non-Abelian parts of the scalar $S=V$ and vector $S=A$ potentials respectively. First we consider a $N_{c}=2$ model with both Abelian and non-Abelian potentials being chosen of the same order of magnitude $||\bar{S}(x)||\sim ||\delta{S}(x)||$, the regime which we shall call generic. We use this model to validate the two representations for the boundary invariant by comparing the results calculated on their basis with the ones calculated from first principles. In complete analogy with the single-channel continuum models [\onlinecite{Miles_21}], we reveal that the boundary invariant exhibits exactly $\nu$ jumps by minus one electron charge unit when the chemical potential is placed in the $\nu^{\text{th}}$ band gap. By analysing the spectral flow of the in-gap edge states, we find that the jumps of $I$ happen exactly at the points where an edge state leaves the occupied part of the spectrum as it is the case in both continuum [\onlinecite{Miles_21}] and tight-binding [\onlinecite{Pletyukhov_2020}, \onlinecite{Pletyukhov_2020_R}, \onlinecite{Muller_2021}] models. Additionally, for the same model, we present the numerical results for the interface invariant and reveal that $Q_{I}$ takes on values in the range $-N_{c}, ..., -1, 0$. 
\par
Further, we analyze the behavior of the boundary invariant as a function of $\frac{||\delta{S}(x)||}{||\bar{S}(x)||}$ in the $N_{c}=3$ model. In particular we cover various regimes ranging from completely degenerate $\frac{||\delta{S}(x)||}{||\bar{S}(x)||}=0$ to the generic $\frac{||\delta{S}(x)||}{||\bar{S}(x)||}\sim 1$ one, and examine the properties of $I$ and the associated spectral flow as the "non-Abelianity" of the potential increases.
\par
Alongside the exact theory, we develop a low-energy approximation of boundary and interface charge invariants covering the universal regime $||\delta{S}(x)||, \ ||\bar{S}(x)||\ll\epsilon_{F}$. In direct analogy with the low-energy theory (LET) one commonly develops in the wave-function language (see for example [\onlinecite{Pletyukhov_2020_prr}, \onlinecite{Piasotski_2021}, \onlinecite{Thakurathi_2018}]), we approximate the bare Green's function of the model by the one written in the basis of slowly varying right and left-moving fields in the vicinity of the chemical potential. With the help of the LET we recover the single-channel results obtained in [\onlinecite{Pletyukhov_2020_prr}] by the wave-function based methods and show that in general, within the LET, the change in the total boundary charge is, up to integer contributions, a linear function of $x_{\varphi}$. Finally, we substantiate our analysis by comparing the results of LET with the exact ones in a particular $N_{c}=2$ model with small potential amplitudes $||\delta{S}(x)||, \  ||\bar{S}(x)||\ll\epsilon_{F}$. 
\par
This paper is organized as follows. In Section \ref{sec: Model} we introduce the class of models considered in this paper and the corresponding single-particle bulk Green's functions. In Section \ref{sec: BGF} the Green's functions for the system with isolated point-like impurity as well as the hard wall boundary (the boundary Green's function) are introduced. In Section \ref{sec: int_charge} the charge on an isolated $\delta$-function impurity (the interface charge) is shown to be a quantized integer quantity. In Sec. \ref{sec: bound_charge} we define the boundary charge in terms of the BGF and study its properties under the shift of the lattice by $x_{\varphi}\in[0, L]$. We define the boundary invariant and derive two equivalent representations for it. In Section \ref{sec: numerics} we provide the numerical validation of the two representations of the boundary invariant and demonstrate the quantization of interface charge. In Sec. \ref{sec: near_degen} we study the boundary invariant as a function of the strength of the non-Abelian part of scalar and vector potentials. In Section \ref{sec: LET} the low-energy approximation of the bulk Green's function and boundary invariant are introduced, field-theoretical version of the surface charge theorem is established, and the analytical results are substantiated with a numerical example. Finally, in Sec. \ref{sec: summary} we state our summary.  
\par
In what follows we set the electron charge $e$ and reduced Plank's constant to unity $e=\hbar=1$.

\section{Model and bulk Green's functions}
\label{sec: Model}

Let us consider a model with the Hamiltonian
\begin{align}
\label{eq: hamiltonian}
H_x = \frac{p^2}{2 m} + \frac{1}{2m} \{ A (x), p\} + V (x),
\end{align}
where $A (x+L) = A (x) = A^{\dagger} (x)$ and $V (x+L) = V(x) = V^{\dagger} (x)$ are $N_c \times N_c$ hermitian matrices, and $p$ is the momentum operator.

The retarded bulk Green's function $G^{(0)} (x,x'; z =\omega+ i \eta ) \equiv G^{(0)} (x,x')$ obeys the equation
\begin{align}
(z - H_x^{\rightarrow})  G^{(0)} (x,x') =G^{(0)} (x,x') (z-H_{x'}^{\leftarrow}) = \delta (x-x'),
\label{Dyson_0}
\end{align}
where the arrows indicate the directions of action of differential operators. Representing
\begin{align}
\nonumber
G^{(0)} (x,x') &= G^{(0)} (\bar{x}+ n L, \bar{x}' + n'L) \\
\label{basis_change_1}
&= \frac{L}{2 \pi} \int_{-\pi/L}^{\pi/L} d k e^{i k (n-n')L} G_k^{(0)}  (\bar{x}, \bar{x}'),
\end{align}
where $\bar{x},\bar{x}' \in [0,L]$, we set up the equations
\begin{align}
(z - H_{\bar{x}}^{\rightarrow})  G_k^{(0)} (\bar{x},\bar{x}') &= \sum_r \delta (\bar{x}-\bar{x}' -r L) e^{i k r L} , \label{eq_G0k}\\
G_k^{(0)} (\bar{x},\bar{x}') (z-H_{\bar{x}'}^{\leftarrow}) &= \sum_r \delta (\bar{x}-\bar{x}' -r L) e^{i k r L} ,
\label{eq_G0kH}
\end{align}
with the boundary conditions
\begin{align}
G_k^{(0)} (L,\bar{x}') = e^{i k L} G_k^{(0)} (0,\bar{x}') , \label{PBC1} \\
G_k^{(0)} (\bar{x},L) = G_k^{(0)} (\bar{x},0) e^{-i k L}. \label{PBC2}
\end{align}
We note the property $[G^{(0)} (x,x') ]^{\dagger} = G^{(0)} (x',x) $, where the hermitian conjugation does not affect $z= \omega + i \eta$. We also used
\begin{align}
\nonumber
\delta (x-x') &= \delta (\bar{x} + (n-1) L-\bar{x}' -(n'-1) L) \\
\nonumber
&= \sum_{r} \delta_{r,n'-n}  \delta (\bar{x} -\bar{x}' -r L) \\
&= \frac{L}{2\pi} \int_{-\pi/L}^{\pi/L} d k e^{i k (n-n') L}  \sum_{r} e^{i k r L}  \delta (\bar{x} -\bar{x}' -r L).
\end{align}

Alternatively, we introduce
\begin{align}
\label{basis_change_2}
G_k^{(0)} (\bar{x},\bar{x}') = e^{i k \bar{x}} \bar{G}_k^{(0)} (\bar{x},\bar{x}')  e^{-i k \bar{x}'},
\end{align}
with $ \bar{G}_k^{(0)} (x,x') $ obeying the equation \eqref{eq_G0k}, in which the momentum operator is shifted $p \to p+k$, and the periodic boundary conditions are imposed
\begin{align}
\bar{G}_k^{(0)} (L,\bar{x}') &= \bar{G}_k^{(0)} (0,\bar{x}') , \\
\bar{G}_k^{(0)} (\bar{x},L) &= \bar{G}_k^{(0)} (\bar{x},0) .
\end{align}
The latter allow us to represent
\begin{align}
\bar{G}_k^{(0)} (\bar{x},\bar{x}') = \frac{1}{L} \sum_{l,l'} \bar{G}_{k,ll'}^{(0)} e^{\frac{2 \pi i}{L} l \bar{x} - \frac{2 \pi i}{L} l'\bar{x}'} .
\label{fourier_repr}
\end{align}
Inserting this expansion into the modified equation \eqref{eq_G0k}, we obtain
\begin{align}
\label{eq: Schodinger_fourier}
\sum_{l''} (z \delta_{ll''} - h_{k,ll''}) \bar{G}_{k,l'' l'}^{(0)} = \delta_{ll'},
\end{align}
where
\begin{align}
h_{k,ll''} = \frac{(\frac{2\pi}{L} l+k)^2}{2m} \delta_{ll''} + \frac{1}{m}\left(\frac{2 \pi}{L} \frac{l+l''}{2} +k \right) A_{l,l''}+ V_{l,l''}, 
\label{Bloch_ham}
\end{align}
and
\begin{align}
V_{l,l''} &\equiv V_{l-l''} = \frac{1}{L} \int_0^L dx V (x) e^{-\frac{2 \pi i}{L} (l-l'') x} , \label{V_Fourier} \\
A_{l,l''} &\equiv A_{l-l''} = \frac{1}{L} \int_0^L dx A (x) e^{-\frac{2 \pi i}{L} (l-l'') x} \label{A_Fourier}
\end{align}
are the Fourier transforms of potential and spin-orbit  matrices.

We remark the following properties of $G^{(0)} (x,x')$: 

1) Jump in the first derivatives across $x = x'$ 
\begin{align}
    G_1^{(0)} (x^{+},x) - G_1^{(0)} (x^{-},x)= G_2^{(0)} (x,x^{+}) - G_2^{(0)} (x,x^{-}) =2 m,
    \label{discont_der}
\end{align}
where $x^{\pm} = x \pm 0^+$, and the indices $1,2$ indicate derivatives with respect to the corresponding arguments; 

2) Jump in the mixed derivative across $x=x'$
\begin{align}
     G_{12}^{(0)} (x^{+},x^{-}) - G_{12}^{(0)} (x^{-},x^{+}) =i 4 m  A (x).
    \label{cont_2nd_der}
\end{align}

\section{Boundary Green's functions}
\label{sec: BGF}

Introducing the impurity potential $\tilde{V}_0 \delta (x)$, we set up the Dyson's equation 
\begin{align}
G (x,x') &= G^{(0)} (x,x') + \tilde{V}_0  G^{(0)} (x,0) G (0,x') \\
&=  G^{(0)} (x,x') + \tilde{V}_0  G (x,0) G^{(0)} (0,x') .
\label{GF_bound}
\end{align}
Establishing first
\begin{align}
G (0,x') =& [1- \tilde{V}_0  G^{(0)} (0,0) ]^{-1} G^{(0)} (0,x') , 
\end{align}
we find the exact solution to the Dyson equation
\begin{align}
G (x,x') =& G^{(0)} (x,x') \nonumber \\
&+ \tilde{V}_0  G^{(0)} (x,0)  [1- \tilde{V}_0  G^{(0)} (0,0) ]^{-1} G^{(0)} (0,x') .
\end{align}
In the limit $\tilde{V}_0 \to \infty$, corresponding to the hard-wall potential, we obtain the BGF
\begin{align}
G (x,x') =& G^{(0)} (x,x')\nonumber \\
&-   G^{(0)} (x,0)  [ G^{(0)} (0,0) ]^{-1} G^{(0)} (0,x') .
\label{Dyson_main}
\end{align}

\section{Interface charge}
\label{sec: int_charge}
To compute the interface charge
\begin{align}
Q_I =&  - \frac{1}{\pi} \text{Im} \int d \omega \Theta (\mu - \omega) \nonumber \\
& \times \int_{-\infty}^{\infty} d x \,   \text{tr} \{   G (x,x) - G^{(0)} (x,x)\},
\end{align}
we use the identity (derived in Appendix \ref{app:int_G0G0})
\begin{align}
 \int_{-\infty}^{\infty} d x \,\,  G^{(0)} (0,x) G^{(0)} (x,0) = -  \frac{\partial G^{(0)}  (0,0)}{\partial \omega}.
 \label{ident_int_G0G0}
\end{align}

The interface charge is then given by
\begin{align}
Q_I =&  -\frac{1}{\pi} \int d \omega \Theta (\mu - \omega) \nonumber \\
 & \times \text{Im } \frac{\partial}{\partial \omega}  \ln \det \left( 1- \tilde{V}_0 G^{(0)} (0,0) \right),
\end{align}
which is an immediate generalization of the corresponding lattice result.

In the limit $ \tilde{V}_0 \to \infty $ we obtain
\begin{align}
Q_I =  -\frac{1}{\pi} \int d \omega \Theta (\mu - \omega) \text{Im } \frac{\partial}{\partial \omega}  \ln \det G^{(0)} (0,0) .
\end{align}

\section{Boundary charge in semi-infinite model}
\label{sec: bound_charge}
\subsection{Definition}

We straightforwardly generalize lattice model expressions to obtain the charge $Q_B = \int_0^{\infty} d x f (x) [\rho (x) - \bar{\rho}]$ accumulated near the (left) boundary in the right semi-infinite model. Hereby $\rho (x)$ is the charge density, $\bar{\rho}$ is the average charge density deep in the bulk (i.e. far away from the boundary), and $f (x)$ is an envelope function, interpolating between the values $1$ close to the boundary and $0$ in the bulk. It can be generally shown [\onlinecite{Pletyukhov_2020}, \onlinecite{Muller_2021}, \onlinecite{Miles_21}] that $Q_B$ can be decomposed as
\begin{align}
& Q_B = Q'_B + Q_P, \label{QB_tot} 
\end{align}
where
\begin{align}
Q'_B = & 
 - \frac{1}{\pi} \text{Im} \int d \omega \Theta (\mu - \omega) \nonumber \\ & \times \int_0^{\infty} d x \, e^{- 0^+ x} \, \text{tr}  \left\{ G (x,x) - G^{(0)} (x,x) \right\}  \label{QB_GmG} \\
=& \frac{1}{\pi} \text{Im} \int d \omega \Theta (\mu - \omega)  \int_0^{\infty} d x \, e^{- 0^+ x} 
\nonumber \\
& \times   \text{tr}  \left\{ [G^{(0)} (0,0)]^{-1} G^{(0)} (0,x) G^{(0)} (x,0) \right\} \label{QB_diff}
\end{align}
represents the sum of the Friedel and the edge state contributions, while
\begin{align}
& Q_P = - \int_0^L d x \frac{x}{L} [\rho^{(0)} (x) - \bar{\rho}] 
\end{align}
is the so called polarization charge expressed in terms of the dipole moment. The latter arises from the charge distribution in the crossover region of the envelope function $f (x)$. Hereby the periodic bulk density and the average bulk density are given by
\begin{align}
\label{eq: density_def}
\rho^{(0)} (x) &=  -\frac{1}{\pi} \text{Im} \int d \omega \Theta (\mu - \omega) \, \text{tr}  \left\{G^{(0)} (x,x) \right\}, \\
\label{eq: density_def_av}
\bar{\rho} &= \frac{1}{L}  \int_0^L d x \rho^{(0)} (x) ,
\end{align}
respectively.

Using the identity (see Appendix \ref{app:int_G0G0}) 
\begin{align}
& 2 m \int_{x_0}^{\infty} d x e^{-0^+ x} G^{(0)} (x_0,x)  G^{(0)} (x,x_0)   \nonumber \\
= & G^{(0)} (x_0,x_0) \frac{\partial  G_1^{(0)} (x_0^+,x_0)}{\partial \omega} -  G_2^{(0)} (x_0,x_0^+) \frac{\partial G^{(0)} (x_0,x_0) }{\partial \omega} \nonumber \\
+&  2 i G^{(0)}(x_0,x_0) A(x_0) \frac{\partial G^{(0)} (x_0,x_0) }{\partial \omega} 
\label{GG_half} \\
= & G^{(0)} (x_0,x_0) \frac{\partial  \mathcal{L} (x_0) }{\partial \omega} G^{(0)} (x_0,x_0)  ,
\label{GG_half_sym}
\end{align}
where 
\begin{align}
    \mathcal{L} (x_0) = [G^{(0)} (x_0,x_0)]^{-1} G_2^{(0)} (x_0,x_0^+) - i A (x_0)
    \label{L_def}
\end{align}
possesses the hermiticity property (see Appendix \ref{app:main_ident})
\begin{align}
\mathcal{L} (x_0) = \mathcal{L}^{\dagger} (x_0),
\label{L_herm}
\end{align}
we cast
\begin{align}
    & \int_{x_0}^{\infty} d x \, e^{- 0^+ x} \, \text{tr}  \left\{ [G^{(0)} (x_0,x_0)]^{-1} G^{(0)} (x_0,x) G^{(0)} (x,x_0) \right\} \nonumber \\
     & = \frac{1}{2m} \text{tr}  \left\{ G^{(0)} (x_0, x_0) \frac{\partial \mathcal{L} (x_0)}{\partial \omega}   \right\}. \label{tr_x0}
\end{align}

Thus, we find that
\begin{align}
Q'_B =  \frac{1}{\pi} \text{Im} \int d \omega \Theta (\mu - \omega)
\text{tr}  \left\{ G^{(0)} (0, 0) \frac{1}{2 m} \frac{\partial \mathcal{L} (0)}{\partial \omega}    \right\}.
\end{align}

\subsection{Properties of $Q_B$ under the system's shift}

Let us shift the right subsystem towards the wall by $x_{\varphi}$ and evaluate an analogue of \eqref{QB_tot} in the shifted system. All quantities referring to the shifted system are denoted by the superscript $(\varphi)$, e.g. $A^{(\varphi)} (x) = A (x+x_{\varphi})$ and $V^{(\varphi)} (x) = V (x+x_{\varphi})$.

By inspecting an equation  for the bulk $G^{(0,\varphi)} (x, x')$, we find that it coincides with $G^{(0)} (x + x_{\varphi}, x'+ x_{\varphi})$.

First we study the change in the quantity $Q_P$
\begin{align}
\Delta Q_P (x_{\varphi}) = Q_P (x_{\varphi}) - Q_P (0) .
\end{align}
Expressing
\begin{align}
\nonumber
 Q_P (x_{\varphi}) =& - \int_0^L d x \frac{x}{L} [\rho^{(0)} (x + x_{\varphi}) - \bar{\rho}] \\
 \label{eq: polarization_def}
 =& - \int_{x_{\varphi}}^{L+x_{\varphi}} d x \frac{x- x_{\varphi}}{L} [\rho^{(0)} (x ) - \bar{\rho}] ,
\end{align}
and noticing that $\rho^{(0)} (x)=\rho^{(0)} (x+L)$ is a periodic function, we derive (see Appendix \ref{app:Delta_QP})
\begin{align}
\Delta Q_P (x_{\varphi}) -  \bar{\rho} x_{\varphi} = - \int_{0}^{x_{\varphi}} d x  \rho^{(0)} (x ) .
\label{dQP}
\end{align}
It is apparent that $\Delta Q_P (L) =0$.

Next, we study $\Delta Q'_B (x_{\varphi}) = Q'_B (x_{\varphi}) - Q'_B (0) $. Considering the Green's function
\begin{align}
    \tilde{G} (x,x') =& G^{(0)} (x,x')\nonumber \\
&-   G^{(0)} (x,x_{\varphi})  [ G^{(0)} (x_{\varphi},x_{\varphi}) ]^{-1} G^{(0)} (x_{\varphi},x') .
\label{Dyson_main_shift}
\end{align}
which is an analogue of \eqref{Dyson_main} with the infinite-strength delta-impurity at $x= x_{\varphi}$ instead of $x=0$, we represent
\begin{align}
 Q'_B (x_{\varphi}) =&
 - \frac{1}{\pi} \text{Im} \int d \omega \Theta (\mu - \omega) \nonumber \\ & \times \int_{x_{\varphi}}^{\infty} d x \, e^{- 0^+ x} \, \text{tr}  \left\{ \tilde{G} (x,x) - G^{(0)} (x,x) \right\}  \\
=&  \frac{1}{\pi} \text{Im} \int d \omega \Theta (\mu - \omega) \int_{x_{\varphi}}^{\infty} d x \, e^{- 0^+ x} 
\nonumber \\
\times &  \text{tr}  \left\{ [G^{(0)} (x_{\varphi},x_{\varphi})]^{-1} G^{(0)} (x_{\varphi},x) G^{(0)} (x,x_{\varphi}) \right\} .
\label{QBshift_diff}
\end{align}
Applying the identity \eqref{tr_x0} to \eqref{QBshift_diff}, we express
\begin{align}
    Q'_B (x_{\varphi}) =& \frac{1}{\pi} \text{Im} \int d \omega \Theta (\mu - \omega) \nonumber \\
\times & \text{tr}  \left\{ G^{(0)} (x_{\varphi}, x_{\varphi}) \frac{1}{2 m} \frac{\partial \mathcal{L} (x_{\varphi})}{\partial \omega}    \right\}.
\end{align}
Since both $G^{(0)} (x_{\varphi,x_{\varphi}})$ and $\mathcal{L} (x_{\varphi})$ are periodic in $x_{\varphi}$, so is $Q'_B (x_{\varphi})$. 
For the difference $\Delta Q'_B (x_{\varphi})= Q'_B (x_{\varphi}) -Q'_B (0)$, which obeys the condition $\Delta Q'_B (L)=0$, we can state the following integral representation
\begin{align}
     \Delta Q'_B (x_{\varphi}) =& \frac{1}{\pi} \text{Im} \int d \omega \Theta (\mu - \omega) \nonumber \\
\times &\int_0^{x_{\varphi}} d x  \frac{d}{dx} \text{tr}  \left\{ G^{(0)} (x,x) \frac{1}{2 m} \frac{\partial \mathcal{L} (x)}{\partial \omega}    \right\}.
\end{align}

\subsection{Representations for the boundary invariant}
\label{sec: reps_I}
We study the boundary invariant
\begin{align}
\label{eq: invariant}
    I (x_{\varphi}) &= \Delta Q_B (x_{\varphi}) - \bar{\rho} x_{\varphi} \\
    &= \Delta Q'_B (x_{\varphi}) -\int_0^{x_{\varphi}} d x \rho^{(0)} (x).
\end{align}
Since $\Delta Q_B (L) = \Delta Q'_B (L)+\Delta Q_P (L)=0$, this invariant possesses the property $I (L) = -\bar{\rho} L$.

Combining the terms, we obtain
\begin{align}
   & I (x_{\varphi})  = \frac{1}{\pi} \text{Im} \int d \omega \Theta (\mu - \omega) \int_0^{x_{\varphi}} d x \nonumber \\
& \times  \text{tr} \left\{ \frac{d}{dx}   \left[G^{(0)} (x,x) \frac{1}{2 m} \frac{\partial \mathcal{L} (x)}{\partial \omega}    \right]  + G^{(0)} (x,x)\right\}.
\label{I_repr1_0}
\end{align}
Another representation reads
\begin{align}
    I (x_{\varphi}) &= - \frac{1}{\pi} \text{Im} \int d \omega \Theta (\mu - \omega) \nonumber \\ & \times \int_{x_{\varphi}}^{\infty} d x \, e^{- 0^+ x} \, \text{tr}  \left\{ \tilde{G} (x,x) - G^{(0)} (x,x) \right\} \nonumber \\
    & + \frac{1}{\pi} \text{Im} \int d \omega \Theta (\mu - \omega) \nonumber \\ & \times \int_0^{\infty} d x \, e^{- 0^+ x} \, \text{tr}  \left\{ G (x,x) - G^{(0)} (x,x) \right\} \nonumber \\
    & + \frac{1}{\pi} \text{Im} \int d \omega \Theta (\mu - \omega) \int_0^{x_{\varphi}}  \, \text{tr} \{ G^{(0)} (x,x)\}.
    \label{I_repr2_0}
\end{align}

\subsubsection{Representation in terms of the winding number}
\label{sec: winding}
Let us first derive an equation for the function $\mathcal{L} (x)$ defined in \eqref{L_def}. Exploiting the relations
\begin{align}
\label{eq: deriv_inver}
    \frac{d}{dx} [G^{(0)} (x,x)]^{-1}  &= - [G^{(0)} (x,x)]^{-1}\nonumber \\
    & \times \frac{d G^{(0)} (x,x)}{dx}  [G^{(0)} (x,x)]^{-1}, \\
    \frac{d G^{(0)} (x,x)}{dx} &=G_1^{(0)} (x^-,x) + G_2^{(0)} (x,x^+),
\end{align}
we establish
\begin{align}
    \frac{d \mathcal{L} (x)}{d x} &=  - [G^{(0)} (x,x)]^{-1} [G_1^{(0)} (x^-,x) + G_2^{(0)} (x,x^+)] \nonumber \\
    & \times  [G^{(0)} (x,x)]^{-1} G_2^{(0)} (x,x^+)\nonumber \\
    &+[G^{(0)} (x,x)]^{-1} [G_{22}^{(0)} (x,x^+) + G_{12}^{(0)} (x^-,x^+)]\nonumber \\
    & -i A' (x).
\end{align}
Using the representation (see Appendix \ref{app:main_ident})
\begin{align}
   G_{12} (x^-,x^+) = G_1^{(0)} (x^-,x)  [G^{(0)} (x,x)]^{-1} G_2^{(0)} (x,x^+),
   \label{G12_ident}
\end{align}
and the equation \eqref{Dyson_0}, we find
\begin{align}
    \frac{d \mathcal{L} (x)}{d x} &= -\mathcal{L}^2 (x) - i  A (x) \mathcal{L} (x)  +i   \mathcal{L} (x) A (x)  \nonumber \\
    & -  A^2 (x) - 2m \left[ z- V(x) \right].
\end{align}

Differentiating it with respect to $\omega$, we also find
\begin{align}
    \frac{d }{d x}  \frac{\partial \mathcal{L} (x)}{\partial \omega} &=  -[ \mathcal{L} (x)+i  A(x)] \frac{\partial \mathcal{L} (x)}{\partial \omega}\nonumber \\
    &-\frac{\partial \mathcal{L} (x)}{\partial \omega} [\mathcal{L}^{\dagger} (x) -i  A (x)]  -2m
\end{align}
and 
\begin{align}
    & \text{tr} \left\{G^{(0)} (x,x) \frac{d }{d x}  \frac{\partial \mathcal{L} (x)}{\partial \omega} \right\} \nonumber \\
    &=  -\text{tr} \left\{ [ G_1^{(0)} (x^+,x) + G_2^{(0)} (x,x^+) ] \frac{\partial \mathcal{L} (x)}{\partial \omega} \right\} \nonumber \\
    &  -2 m \, \text{tr} \{G^{(0)} (x,x)\} .
\end{align}
From the last relation we obtain
\begin{align}
    & \text{tr} \left\{ \frac{d}{dx}   \left[G^{(0)} (x,x) \frac{1}{2 m} \frac{\partial \mathcal{L} (x)}{\partial \omega}    \right]  + G^{(0)} (x,x)\right\} \nonumber \\
    &= -\text{tr} \left\{ \frac{\partial \mathcal{L} (x)}{\partial \omega} \right\} .
\end{align}
Substituting it in \eqref{I_repr1_0}, we reveal 
\begin{align}
      & I (x_{\varphi})  = -\frac{1}{\pi} \text{Im} \int d \omega \Theta (\mu - \omega) \frac{\partial}{\partial \omega}\int_0^{x_{\varphi}} d x \,  \text{tr} \left\{   \mathcal{L} (x)\right\}.
\label{I_repr1_1}
\end{align}

Defining the matrix
\begin{align}
    \mathcal{U}(x)=\text{P}\!\exp \left\{ \int_0^x d x' \mathcal{L} (x') \right\},
\end{align}
which satisfies the equation
\begin{align}
    \frac{d \mathcal{U} (x)}{d x} = \mathcal{L} (x)  \mathcal{U} (x), \quad \mathcal{U} (0)=1,
    \label{eq_U}
\end{align}
and using the Jacobi's formula
\begin{align}
\text{tr} \left\{   \mathcal{L} (x)\right\} = \text{tr} \, \left\{ \frac{d \mathcal{U} (x)}{d x} \mathcal{U}^{-1} (x) \right\} = \frac{d}{d x} \ln \det \mathcal{U} (x),
\end{align}
we express the boundary invariant as the winding number of $\det \, \mathcal{U} (x_{\varphi})$: 
\begin{align}
      & I (x_{\varphi})  = -\frac{1}{\pi} \text{Im} \int d \omega \Theta (\mu - \omega) \frac{\partial}{\partial \omega} \ln \det  \mathcal{U} (x_{\varphi}).
\label{I_repr1_2}
\end{align}

When the chemical potential is located above the $\nu^{\text{th}}$ band the average density in the bulk amounts to $\bar\rho=\frac{\nu}{L}$ (see Appendix \ref{ap: av_des}), and by \eqref{eq: invariant} it follows $I (L)=-\bar\rho L=-\nu$. This defines the following topological invariant associated with the filling factor $\nu$
\begin{align}
      & \mathcal{I}_{\nu}=-I(L) = \frac{1}{\pi} \text{Im} \int d \omega \Theta (\mu - \omega) \frac{\partial}{\partial \omega} \ln \det  \mathcal{W} =\nu,
\label{I_repr1_2}
\end{align}
where $\mathcal{W}$ is given by
\begin{align}
    \mathcal{W}=\mathcal{U}(L)=\text{P}\!\exp \left\{ \int_0^L d x \mathcal{L}(x) \right\}.
\end{align}
Let us note that $\mathcal{L} (x)$ may be expressed in terms of the bulk quantities solely. On the other hand, $\mathcal{I}_{\nu}$ provides the information on the number of edge states sitting in the gap above the band $\nu$. Indeed, since $I (x_{\varphi})$ drops by unity each time an edge state escapes the occupied part of the spectrum, each time the edge state escapes it never returns back as the edge state dispersion always connects the valence and conduction bands together [\onlinecite{Pletyukhov_2020}, \onlinecite{Muller_2021}, \onlinecite{Miles_21}] (see also Sec. \ref{sec: numerics} and \ref{sec: near_degen}), and the in-gap edge state dispersion relation is periodic in $x_{\varphi}$, $\mathcal{I}_{\nu}=\nu$ is precisely equal to the number of edge states to be found over one $x_{\varphi}$-cycle in the gap above the band $\nu$. We thus see that $\mathcal{I}_{\nu}$ underpins a new form of \textit{bulk-boundary correspondence} (BBC) in multichannel one-dimensional models. As it is emphasised in Sec. \ref{sec: near_degen} the conventional form of BBC in the sense of Hatsugai [\onlinecite{Hatsugai_1993}] cannot be established in multichannel models, due to the non-vanishing $x_{\varphi}$ derivative of the edge state dispersion relation at the band touching points.  

\subsubsection{Representation in terms of bound states}
\label{sec: bound_states}
Let us introduce the Green's function $G^{(2)} (x,x')$ for the model with two infinite-strength delta-impurities at $x=0$ and $x=x_{\varphi}$. It obeys the relations
\begin{align}
    G^{(2)} (x,x') &= \tilde{G} (x,x'), \quad x,x' > x_{\varphi}, \\
    G^{(2)} (x,x') &= G (x,x'), \quad x,x'<0,
\end{align}
and for $0< x,x'< x_{\varphi}$ it satisfies the Schr{\"o}dinger equation with open boundary conditions. This observation allows us to rewrite \eqref{I_repr2_0} in the form
\begin{align}
& I (x_{\varphi})= \frac{1}{\pi} \text{Im} \int d \omega \Theta (\mu - \omega)  \int_0^{x_{\varphi}} d x \,   \text{tr} \{   G^{(2)} (x,x)  \}
 \nonumber \\
&- \frac{1}{\pi} \text{Im} \int d \omega \Theta (\mu - \omega)  \int_{-\infty}^{\infty} d x \,  \text{tr} \{  G^{(2)} (x,x) - G (x,x)  \} .
\end{align}
The first contribution is interpreted as $-N_b$, where $N_b$ is a number of bound states in the isolated cavity $0<x<x_{\varphi}$, which lie below the chemical potential $\mu$. This follows from the fact that $G^{(2)}$ is the exact Green's function for the cavity of size $x_{\varphi}$ with open boundary conditions. In the Lehmann representation we get
\begin{align}
\nonumber
    &\frac{1}{\pi} \text{Im} \int d \omega \Theta (\mu - \omega)  \int_0^{x_{\varphi}} d x \,   \text{tr} \{   G^{(2)} (x,x)  \}\\
    \nonumber
    &=\frac{1}{\pi} \text{Im} \int d \omega \Theta (\mu - \omega)  \int_0^{x_{\varphi}} d x \, \\
    \nonumber
    &\times\sum_{\beta=1}^{\infty}  \frac{1}{\omega-\epsilon_{\beta}+i\eta}\text{tr} \{ \psi_{\beta}(x)\psi_{\beta}^{\dagger}(x)\}\\
    \nonumber
    &=  - \,  \sum_{\epsilon_{\beta}<\mu}\ \int_0^{x_{\varphi}} d x\text{tr} \{  \psi_{\beta}(x)\psi_{\beta}^{\dagger}(x)\}\\
    &=  - \,  \sum_{\epsilon_{\beta}<\mu}\ \int_0^{x_{\varphi}} d x|\psi_{\beta}(x)|^{2}=-N_{b},
\end{align}
where $\psi_{\beta}(x)$ are the exact eigenfunctions of the Hamiltonian \eqref{eq: hamiltonian} and $\epsilon_{\beta}$ are the corresponding eigenvalues for the given boundary conditions.

To evaluate the second term, we note the equation for $G^{(2)} (x,x')$
\begin{align}
    G^{(2)} (x,x') &= G (x,x') \nonumber \\
    &- G (x,x_{\varphi}) [G (x_{\varphi},x_{\varphi})]^{-1} G (x_{\varphi},x').
\end{align} 
Using the analog of \eqref{ident_int_G0G0} for $G(x,x')$ (see Appendix \ref{app:int_G0G0})
\begin{align}
     \int_{-\infty}^{\infty} d x \,  G (x_{\varphi},x) G (x,x_{\varphi}) =  - \frac{\partial G (x_{\varphi}, x_{\varphi})}{\partial \omega},
    \label{ident_int_GG}
\end{align}
we obtain
\begin{align}
I (x_{\varphi}) =-N_b -\frac{1}{\pi} \text{Im} \int d \omega \Theta (\mu - \omega)  \frac{\partial }{\partial \omega} \ln \det G (x_{\varphi}, x_{\varphi}) ,
\label{I_repr2_1}
\end{align}
i.e. the boundary invariant can be expressed in terms of the bound states and the winding number of $\det G (x_{\varphi}, x_{\varphi})$.

The comparison of \eqref{I_repr1_2} and \eqref{I_repr2_1} also provides the formula for $N_b$:
\begin{align}
\label{eq: num_bound_states}
    N_b =-\frac{1}{\pi} \text{Im} \int d \omega \Theta (\mu - \omega)  \frac{\partial }{\partial \omega} \ln \det [G (x_{\varphi}, x_{\varphi}) \mathcal{U}^{-1} (x_{\varphi})].
\end{align}

\section{Numerical results. Validation of boundary and interface invariants}
\label{sec: numerics}
\begin{figure*}[t]
                \includegraphics[width=2.0\columnwidth]{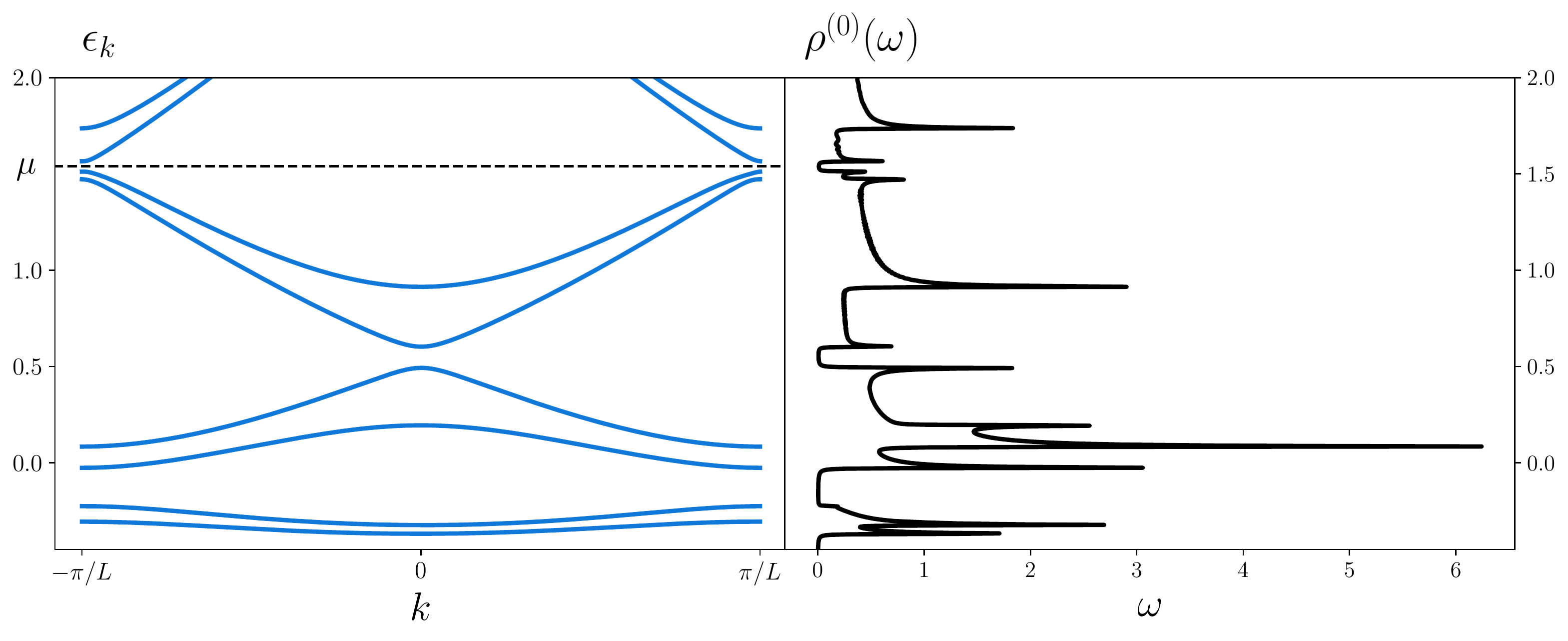}
                \caption{The bulk energy spectrum of the model used in the numerical calculations throughout this section. Left panel: band structure corresponding to the model defined by the non-Abelian gauge potentials (\ref{eq:A2}), (\ref{eq:V2}) with $L=5 \ \text{a.u}$. Right panel: the corresponding bulk local spectral density at the origin of position space $\rho^{(0)}(\omega)=\rho^{(0)}(\omega, x=0)=-\frac{1}{\pi}\text{Im}G^{(0)}(0, 0)$ (for numerical reasons $\eta=10^{-3}$ was chosen in the calculation of spectral density). Throughout the numerical calculations we assume that the chemical potential $\mu$ is located in the gap above the sixth energy band, which we schematically illustrate by the black dashed line.}
                \label{fig: spectrum}
\end{figure*}

In order to showcase the above developed theory we study a particular $N_{c}=2$ system defined by the following potential profiles
\begin{align}
\nonumber
    A(x)=&0.6\sigma_{0}\sin(qx)+0.7\sigma_{3}\sin(3qx)\\
    \label{eq:A2}
    +&0.4[\cos(\pi/4)\sigma_{1}+\sin(\pi/7)\sigma_{2}]\cos(4qx),\\
    \nonumber
    V(x)=&0.3\sigma_{0}\cos(qx)+0.3\sigma_{1}\sin(2qx)\\
    \label{eq:V2}
    -&0.4\sigma_{2}\cos(2qx),
\end{align}
where $\sigma_{0}$ is the $2\times2$ identity matrix, $\sigma_{j}, \ (j=1, 2, 3)$ are the usual Pauli matrices,  $q=\frac{2\pi}{L}$, and we set $L=5\ \text{a}.\text{u}.$ (throughout the calculations we adopt the Hartree's atomic units, that is we additionally set the electron mass to unity $m=1$). The energy dispersion of the bulk system is shown in Figure \ref{fig: spectrum}. In the following we assume that the chemical potential of the system is located in the band gap above the sixth band $\max_{k}\epsilon_{6, k}<\mu<\min_{k}\epsilon_{7, k}$.

\subsection{Calculation of winding numbers and bulk position space Green's functions}
\label{sec: numerics_gen}
\begin{figure}[t]                \includegraphics[width=0.92\columnwidth]{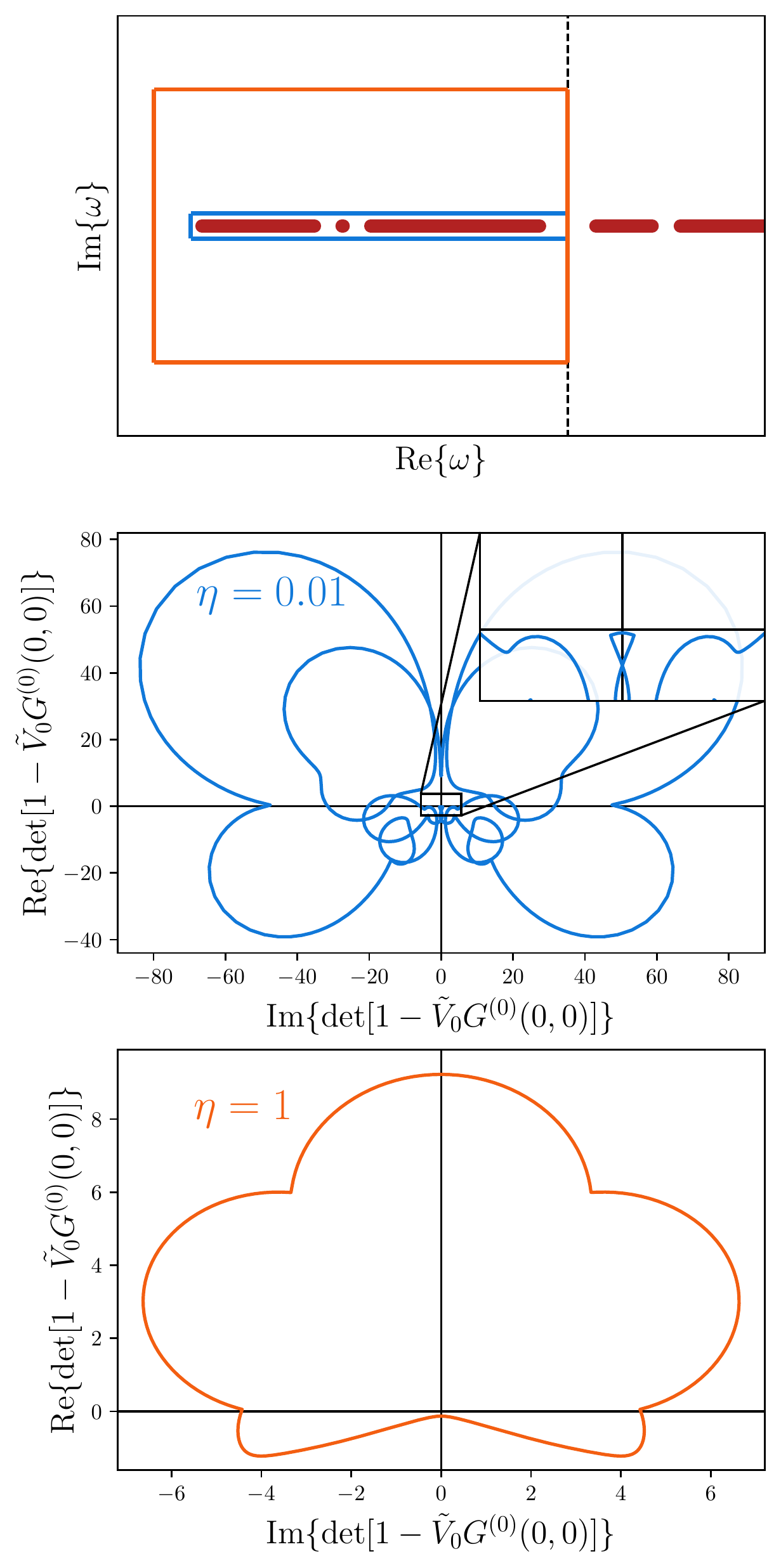}
                \caption{Top panel: Schematic illustration of two contours in the the complex $\omega$-plane encompassing the non-analytical features of the $K$-function on the real axis. Here blue and orange contours correspond to the small and large values of $\eta, \ B$ respectively. Second and third panels show a comparison between the mappings of the small and large $\eta$ rectangular contours by the function $K=\det[1-\tilde{V}_{0}G^{(0)}(0, 0)]$. Despite of the smaller $\eta$ contour having much richer structure than its large $\eta$ counterpart, we see that the value of winding number ($-1$) is unaffected by contour deformations as is suggested in main text. }
                \label{fig: contour_eta}
\end{figure}

We are interested in evaluation of the winding numbers of the form
\begin{align}
    \text{wn}[K]=-\int\frac{d\omega}{\pi}\Theta(\mu-\omega)\text{Im}\frac{\partial}{\partial\omega}\ln{K}(\omega),
\end{align}
for some function $K(\omega)$. In order to evaluate $\text{wn}[K]$ we rewrite the above integral as
\begin{align}
\label{eq: contour_representation}
    \text{wn}[K]=-\lcirclerightint_{C}\frac{d\omega}{2\pi i}\frac{\partial}{\partial\omega}\ln{K}(\omega),
\end{align}
where $C$ is a clockwise contour in the complex frequency plane surrounding the occupied part of the spectrum residing on the real axis. For practical purposes we consider a rectangular contour having the width $2\eta$ in the imaginary direction and ranging from $B$ to $\mu$ in the real one. Here $\mu$ is the chemical potential and $B$ is some energy lying below the bottom of the lowest band. Let us also note that nonetheless retarded and advanced Green's functions are defined in the limit $\eta\rightarrow 0^{+}$, in practical calculations of winding numbers, $\eta$ may be chosen arbitrarily (and it is numerically beneficial to choose larger values of $\eta$) since the winding number, being a member of the homotopy class of a unit circle, simply measures the degree of the mapping $S^{1}\rightarrow S^{1}$ and cannot be affected by \textit{continuous} contour deformations (see Fig. \ref{fig: contour_eta}). Numerically, the contour integral in (\ref{eq: contour_representation}) is calculated as [\onlinecite{Muller_2021}]
\begin{align}
    \lcirclerightint_{C}\frac{d\omega}{2\pi i}\frac{\partial}{\partial\omega}\ln{K}(\omega)\approx\sum_{n=1}^{n_{c}}\Delta_{n}^{K}, \ n_{c}\gg1,
\end{align}
where 
\begin{align}
\Delta_{n}^{K}&=\frac{1}{2\pi}\begin{cases}\frac{Y^{K}_{n}\delta{X}^{K}_{n+1, n}-X^{K}_{n}\delta{Y}^{K}_{n+1, n}}{(R^{K}_{n})^{2}}, \quad n\leq n_{c}-1, \\ \frac{Y^{K}_{n}\delta{X}^{K}_{1, n_{c}}-X^{K}_{n}\delta{Y}^{K}_{1, n_{c}}}{(R^{K}_{n_{c}})^{2}}, \quad n= n_{c}, \end{cases}\\
X^{K}_{n}&=\text{Re}[K(\omega_{n})], \quad Y^{K}_{n}=\text{Im}[K(\omega_{n})], \\
R^{K}_{n}&=\sqrt{(X^{K}_{n})^{2}+(Y^{K}_{n})^{2}},\\
\delta{X^{K}_{a, b}}&=X^{K}_{a}-X^{K}_{b}, \quad \delta{Y^{K}_{a, b}}=Y^{K}_{a}-Y^{K}_{b}.
\end{align}
Here $\{\omega_{n}\}_{n=1, ..., n_{c}}$ is the contour-ordered set of points on the contour $C$. Note that for small values of $\eta$, in the spirit of [\onlinecite{Muller_2021}], one has to employ the adaptive numerical algorithm for the determination of points $\{\omega_{n}\}_{n=1, ..., n_{c}}$ for any given calculation of the winding number. Using the above mentioned analytical continuation trick (choosing larger values of $\eta$) one automatically avoids these complications since the said $K$-function is only needed far away from the non-analytical features of $G^{(0)}$. This allows one to evaluate all of the ingredients involved in the calculation ($G^{(0)}$, $G^{(0)}_{1}$, $G^{(0)}_{2}$) once on the fixed grid of $\omega$-points. Despite this drastic simplification, we stress that as a rule of thumb one has to choose a much denser set of $\omega$-points in the vicinity of the chemical potential $\omega\in\{\mu+iy| y\in[\eta, -\eta]\}$ than on the rest of the contour $\omega\in C\setminus\{\mu+iy| y\in[\eta, -\eta]\}$ (in fact, for $\eta, \ B=\mathcal{O}(1)$ the points on the rest of the contour may be chosen rather sparsely). This very observation served as our major motivation for the development of the low-energy theory presented in Section \ref{sec: LET}.
\par
As we have seen in the Sections \ref{sec: int_charge} and \ref{sec: bound_charge}, the relevant invariants may be expressed through the functionals of the \textit{bulk} position-space Green's function. For these purposes we employ the series of basis transformations (\ref{basis_change_1}), (\ref{basis_change_2}), (\ref{fourier_repr}), and search for the Fourier components of the bulk Green's function $\bar{G}^{(0)}_{k, l, l'}$. In order to numerically solve the equation (\ref{eq: Schodinger_fourier}) for $\bar{G}^{(0)}_{k, l, l'}$ we introduce a cutoff $M<\infty$ in the space of Fourier modes (labeled by $l, l'$) and replace the infinite range Fourier summations by the corresponding cutoff ones as
\begin{align}
\sum_{l=-\infty}^{\infty}f_{l}\approx\sum_{l=-M}^{M}f_{l},
\end{align}
for a given sequence $f_{l}$. The value of the cutoff has to be chosen in the way that leads to the convergence of the corresponding observables. Within this approximation we write (\ref{eq: Schodinger_fourier}) as 
\begin{align}
(z\hat{1}-\hat{h}_{k})\hat{\bar{G}}^{(0)}_{k}\approx\hat{1},
\end{align}
where $\hat{1}$ is the $M\cdot N_{c}\times M\cdot N_{c}$ identity matrix, $\hat{\bar{G}}^{(0)}_{k}$ is the $M\times M$ block matrix composed of $N_{c}\times N_{c}$ blocks $\bar{G}^{(0)}_{k, l, l'}$, and $z\in\{\omega_{n}\}_{n=1, ..., n_{c}}$. It thus follows that 
\begin{align}
\bar{G}^{(0)}_{k, l, l'}\approx(z\hat{1}-\hat{h}_{k})^{-1}_{l, l'}.
\end{align}
And hence
\begin{align}
\bar{G}^{(0)}_{k}(\bar{x}, \bar{x}')\approx\frac{1}{L}\sum_{l, l'=-M}^{M}(z\hat{1}-\hat{h}_{k})^{-1}_{l, l'}e^{\frac{2\pi i}{L}(l\bar{x}-l'\bar{x}')}.
\end{align}
In order to perform the integral in (\ref{basis_change_1}) we discretize the first Brillouin zone by slicing it into $N_{k}\gg1$ segments of width $\delta_{k}=\frac{2\pi}{(N_{k}-1)L}$ and replace the integration with a sum of $N_{k}$ evenly spaced momenta $k_{m}=-\frac{\pi}{L}+\delta_{k}m, \quad m=0,\ ...,\ N_{k}-1$, so that
\begin{align}
\int_{-\pi/L}^{\pi/L}dkf(k)\approx\delta_{k}\sum_{k_{m}}f(k_{m}).
\end{align}
Taking into account the transformation (\ref{basis_change_2}) it finally follows
\begin{align}
\nonumber
G^{(0)}(x, x')\approx&\frac{\delta_{k}}{2\pi}\sum_{k_{m}}e^{ik_{m}(\bar{x}+nL)}e^{-ik_{m}(\bar{x}'+n'L)}\\
\label{eq: GF_numerics1}
\times&\sum_{l, l'=-M}^{M}(z\hat{1}-\hat{h}_{k_{m}})^{-1}_{l, l'}e^{\frac{2\pi i}{L}(l\bar{x}-l'\bar{x}')}.
\end{align}
We note that in practice, the inverse of $z\hat{1}-\hat{h}_{k}$ is numerically ill defined for $z$ in the vicinity of the real axis. Indeed, since the above matrix is normal $[z\hat{1}-\hat{h}_{k}, (z\hat{1}-\hat{h}_{k})^{\dagger}]=0$, its conditional number may be written as [\onlinecite{Cheney}]
\begin{align}
\kappa(z\hat{1}-\hat{h}_{k})=\frac{\max(|z-\text{spec}\{\hat{h}_{k}\}|)}{\min(|z-\text{spec}\{\hat{h}_{k}\}|)},
\end{align}
and since $\min(|z-\text{spec}\{\hat{h}_{k}\}|)$ takes arbitrarily small values as one approaches the spectrum of $\hat{h}_{k}$, we conclude that $\kappa$ may take on arbitrarily large values, so that the numerical inverse ceases to exist (independently of the numerical inversion algorithm). To avoid this complication, we perform numerical diagonalization of $\hat{h}_{k}$ and rewrite the inverse as
\begin{align}
(z\hat{1}-\hat{h}_{k})^{-1}=\sum_{\alpha=1}^{N_{\alpha}}\frac{1}{z-\epsilon_{\alpha, k}}\ket{\alpha, k}\bra{\alpha, k}, 
\end{align}
where $\epsilon_{\alpha, k}, \ \ket{\alpha, k}$ are the eigenvalues and the normalized eigenstates of $\hat{h}_{k}$ labeled by the band index $\alpha=1, ..., N_{\alpha}=N_{c}\cdot(2M+1)$. With this in hands we rewrite the equation (\ref{eq: GF_numerics1}) as 
\begin{align}
\nonumber
G^{(0)}(x, x')\approx&\frac{\delta_{k}}{2\pi}\sum_{k_{m}}e^{ik_{m}(\bar{x}+nL)}e^{-ik_{m}(\bar{x}'+n'L)}\\
\label{eq: GF_numerics2}
\times&\sum_{\alpha=1}^{N_{\alpha}}\frac{1}{z-\epsilon_{\alpha, k_{m}}}F_{\alpha, m}(\bar{x}, \bar{x}'),
\end{align}
where we have defined 
\begin{align}
\label{eq: F-function}
F_{\alpha, m}(\bar{x}, \bar{x}')=\sum_{l, l'=-M}^{M}[\ket{\alpha, k_{m}}\bra{\alpha, k_{m}}]_{l, l'}e^{\frac{2\pi i}{L}(l\bar{x}-l'\bar{x}')}.
\end{align}
Note that the above representation is also beneficial from the numerical perspective since the auxiliary function $F_{\alpha, m}(\bar{x}, \bar{x}')$ requires a one-off numerical evaluation once $M, \ N_{k}$ and the position-space grid have been established. Analogous representations hold for the derivatives of the bare Green's function $G^{(0)}_{1}(x, x')$ and $G^{(0)}_{2}(x, x')$.

\subsection{First principles calculation of boundary invariant}
\label{sec: First_princ}
To test the validity of the representations of the boundary invariant introduced in Sections \ref{sec: winding} and \ref{sec: bound_states}, we compare the invariant computed based on equations (\ref{I_repr1_2}) and (\ref{I_repr2_1}) with the one computed from the first principles. As opposed to our analysis of boundary invariant in the multichannel tight-binding models carried out in [\onlinecite{Muller_2021}], we find that the calculation of the invariant based on the exact numerical diagonalization of the Hamiltonian (\ref{eq: hamiltonian}) in the basis of standing waves is unfeasible due to the size of the dimension of its matrix representation required for the convergence of the results. To this end, we again employ the method of boundary Green's functions. In particular, we make use of the formula (\ref{I_repr1_0}), which is derived from the basic definition of the boundary charge by simple position space integration. 
\par
As one may infer, handling the formula (\ref{I_repr1_0}) requires (in addition to the calculation of the bulk Green's functions and its spatial derivatives) the knowledge of the frequency derivative of the $\mathcal{L}(x)$ function defined by (\ref{L_def}). Exploiting the matrix identity (\ref{eq: deriv_inver}) we find
\begin{align}
\nonumber
    &G^{(0)}(x, x)\frac{\partial}{\partial\omega}\mathcal{L}(x)=\frac{\partial G_{2}^{(0)}(x, x^{+})}{\partial\omega}\\
&-\frac{\partial G^{(0)}(x, x)}{\partial\omega}[G^{(0)}(x, x)]^{-1}G_{2}^{(0)}(x, x^{+}).
\end{align}
Here we again find the representation (\ref{eq: GF_numerics2}) beneficial since it allows one to avoid the finite-difference calculation of $\omega$ derivative and simply write it as
\begin{align}
\nonumber
\frac{\partial}{\partial\omega}G^{(0)}(x, x')\approx&-\frac{\delta_{k}}{2\pi}\sum_{k_{m}}e^{ik_{m}(\bar{x}+nL)}e^{-ik_{m}(\bar{x}'+n'L)}\\
\label{eq: GF_numerics_der_omega}
\times&\sum_{\alpha=1}^{N_{\alpha}}\frac{1}{(z-\epsilon_{\alpha, k_{m}})^{2}}F_{\alpha, m}(\bar{x}, \bar{x}').
\end{align}
Again, an analogous representation holds for $\frac{\partial}{\partial\omega}G^{(0)}_{2}(x, x')$. Evaluation of the integral in (\ref{I_repr1_0}) is again eased by rewriting it as a contour integral in the spirit of (\ref{eq: contour_representation}) with the following analytical continuation. 

\subsection{Results}
\label{sec: numerics_results}

\subsubsection{Interface charge}
\begin{figure*}
                \includegraphics[scale=0.33]{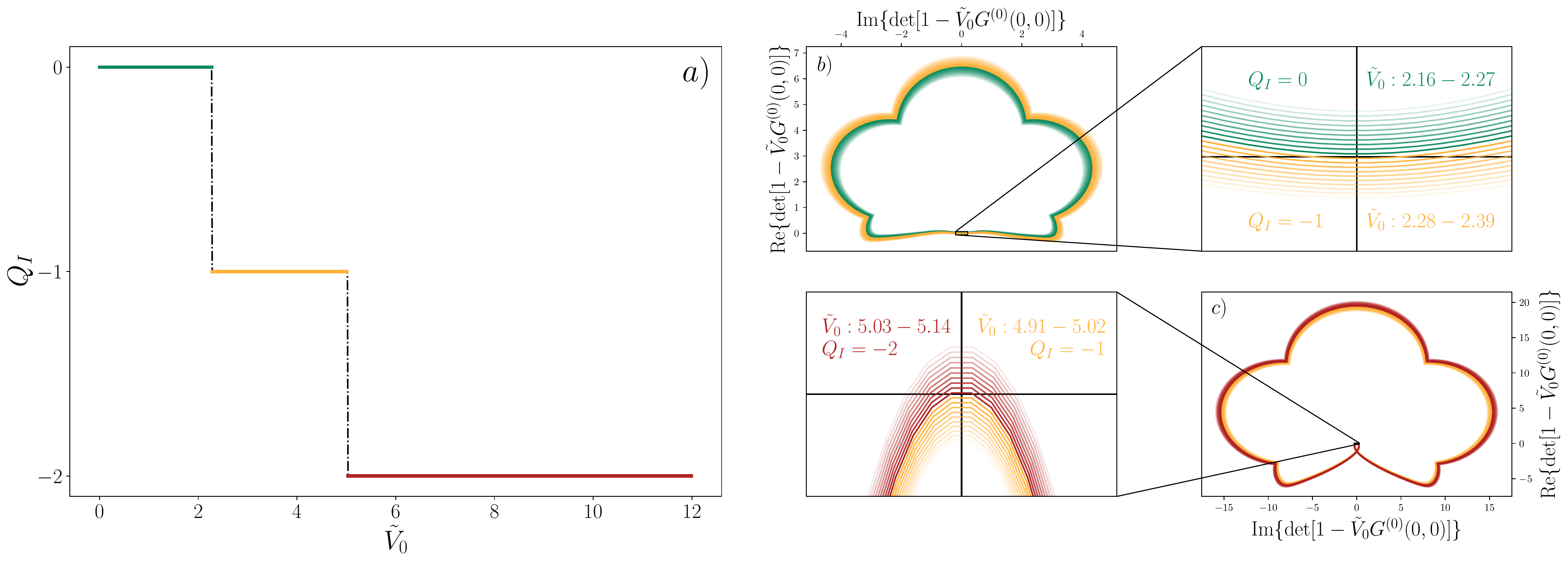}
                \caption{Panel a) shows the interface charge as a function of the impurity strength $\tilde{V}_{0}$ (here $\tilde{V}_{0}$ is measured in Hartree units) for the model defined by potentials (\ref{eq:A2}), (\ref{eq:V2}). As one can see, the charge on the impurity is restricted to the values $-2 \, (=-N_{c}),\ ...,\ 0$ as it was the case in tight-binding description [\onlinecite{Muller_2021}]. Panels b) and c) demonstrate how the contour defined by the $K$-function $\det[1-\tilde{V}_{0}G^{(0)}(0, 0)]$ crosses the point $0+i0$ when the interface charge drops as $0\rightarrow-1$ and $-1\rightarrow-2$ respectively. }
                \label{fig: QI}
\end{figure*}
In this subsection we start our discussion by considering the interface charge invariant as a function of the impurity strength $Q_{I}(\tilde{V}_{0})$. Numerical results for the interface charge invariant are presented in panel a) of Fig. \ref{fig: QI} for $10^{3}$ values of $\tilde{V}_{0}$ in the range $[0, 12]$. The chemical potential was chosen to lie slightly above the sixth energy band $\mu=\max_{k}\epsilon_{6, k}+10^{-2}$. As opposed to the single-channel case where the charge on an isolated impurity is restricted to the values $-1, 0$ [\onlinecite{Pletyukhov_2020_prr}, \onlinecite{Miles_21}], we observe that the interface charge in the multichannel system takes on values $-2 \, (=-N_c), \ ..., \ 0$. 
\par
The jumps of the interface charge occur whenever an edge state pole defined by the condition $\det[1-\tilde{V}_{0}G^{(0)}(0, 0)]\big{|}_{\omega\in\mathbb{R}}=0$ crosses the chemical potential from below, that is a unit of electron charge is carried away from the system in the process of spectral flow. In panels b) and c) we demonstrate this process by studying the flow of the contours resulting from the mapping of the rectangular contour defined in Sec. \ref{sec: numerics_gen} by the $K$-function $K(\omega)=\det[1-\tilde{V}_{0}G^{(0)}(0, 0)]$ as a function of impurity strength $\tilde{V}_{0}$. Once the first bound state leaves the system, the $K$-contour crosses the origin of the complex plane $\omega=0+i0$ leading to the winding number of $-1$. As the value of the impurity strength increases even further, the second edge state escapes the occupied part of the spectrum resulting in the second drop of $Q_{I}$ by unity which is reflected by additional knotting of the contour around $\omega=0+i0$.

\subsubsection{Boundary invariant}
\begin{figure}
                \includegraphics[scale=0.6]{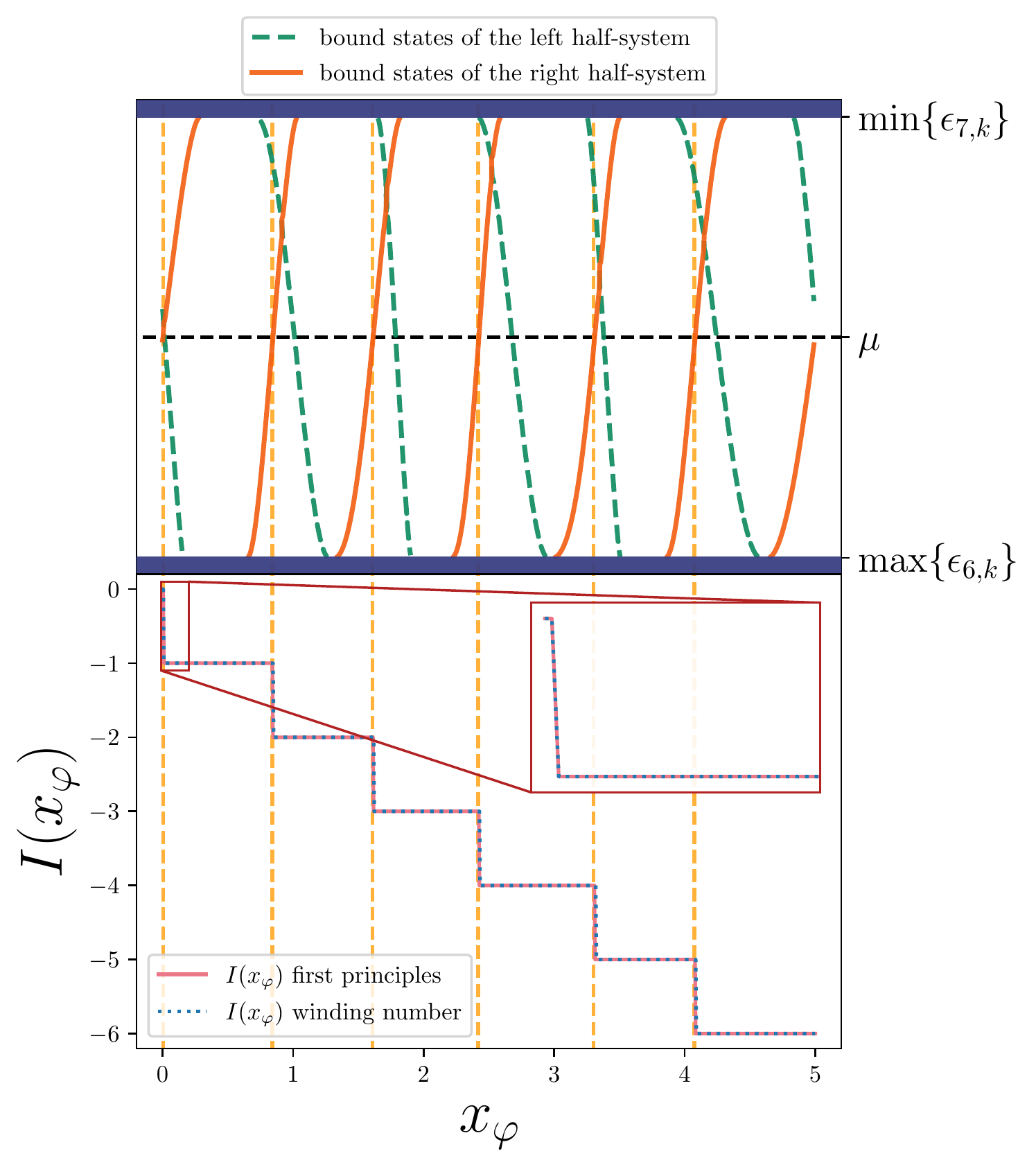}
                \caption{Top panel: the spectral flow of the in-gap edge states as a function of the shift variable $x_{\varphi}$. Here the orange solid lines indicate the bound state poles of the right half-system and the green dashed lines indicate those corresponding to the left half-system. Position of the chemical potential (indicated by the dashed black line) is chosen to be straight in the middle of the band gap between the sixth and seventh bands. Bottom panel shows the boundary invariant $I(x_{\varphi})=\Delta Q_{B}(x_{\varphi})-\bar\rho x_{\varphi}$ as defined in (\ref{eq: invariant}). Invariant calculated from the first principles (utilizing the equation (\ref{I_repr1_0})) is shown in pink, whereas the boundary invariant calculated on the basis of the winding number expression (\ref{I_repr1_2}) is depicted in blue dots. As one may notice the value of the invariant drops by $1$ each time the bound state pole crosses the chemical potential from below.}
                \label{fig: I_winding}
\end{figure}
Now let us turn our attention to the boundary invariant $I(x_{\varphi})$. Boundary invariant (BI) computed on the basis of the winding number representation (\ref{I_repr1_2}) alongside the one computed from the first-principles (see Sec. \ref{sec: First_princ}) is shown in the bottom panel of Fig. \ref{fig: I_winding}. In the calculations of $I(x_{\varphi})$ the chemical potential was chosen to lie straight in the middle of the relevant gap $\mu=\frac{1}{2}(\max_{k}\epsilon_{6, k}+\min_{k}\epsilon_{7, k})$ and the position-space grid consists of $10^{3}$ evenly spaced points in the range $[0, L]=[0, 5]$. First, we note that BIs calculated within two approaches are in exquisite agreement with one another, thus justifying the winding number representation of Section \ref{sec: winding}. In analogy with the single-channel continuum models [\onlinecite{Miles_21}], BI shows exactly $\nu=6$ downward jumps by a unit of electron charge for the filling factor $\nu$ leading to the properties $I(0)=0, \ I(L)=-\bar\rho L=-\frac{\nu}{L}L=-\nu=-6$ demonstrated analytically in Sec. \ref{sec: reps_I}. 
\par
Again the mechanism leading to the abrupt reduction of the charge is the spectral flow of the edge state poles inside the gap hosting the chemical potential. The pole positions as a function of the shift parameter $x_{\varphi}$, found as the real solutions to $\det[G^{(0)}(x_{\varphi}, x_{\varphi})]=0$, are plotted in the top panel of Fig. \ref{fig: I_winding}.   We note that there are precisely $12$ solutions to the pole equation in the interval $x_{\varphi}\in[0, L]$. Six of these, plotted in dashed green lines, correspond to the left half-system, whereas the other six, depicted in solid orange lines, are the ones relevant for us, those corresponding to the right half-system. As one may infer from Fig. \ref{fig: I_winding}, the downward steps of $I(x_{\varphi})$ happen exactly at the points where bound states leave the system, as is indicated by vertical yellow lines.
\par
Let us note that the entire bulk band structure of any multichannel continuum model is invariant under continuous shifts of the lattice by $x_{\varphi}$ (as it is the case in single-channel continuum models [\onlinecite{Miles_21}]) as opposed to the tight-binding models examined in [\onlinecite{Muller_2021}]. This fact may be trivially seen by a simple coordinate shift in the Schrodinger equation for the bulk eigenstates of the system. Another similarity to the single-channel continuum theory is the presence of exactly $\nu=6$ poles in the bandgap above the $\nu^{\text{th}}$ band, an interesting difference though lies in the discontinuity of the edge state dispersion relation, that is, the poles of the left half-system do not continuously transform into the ones of the right half in the process of the spectral flow.  
\begin{figure}
                \includegraphics[scale=0.5]{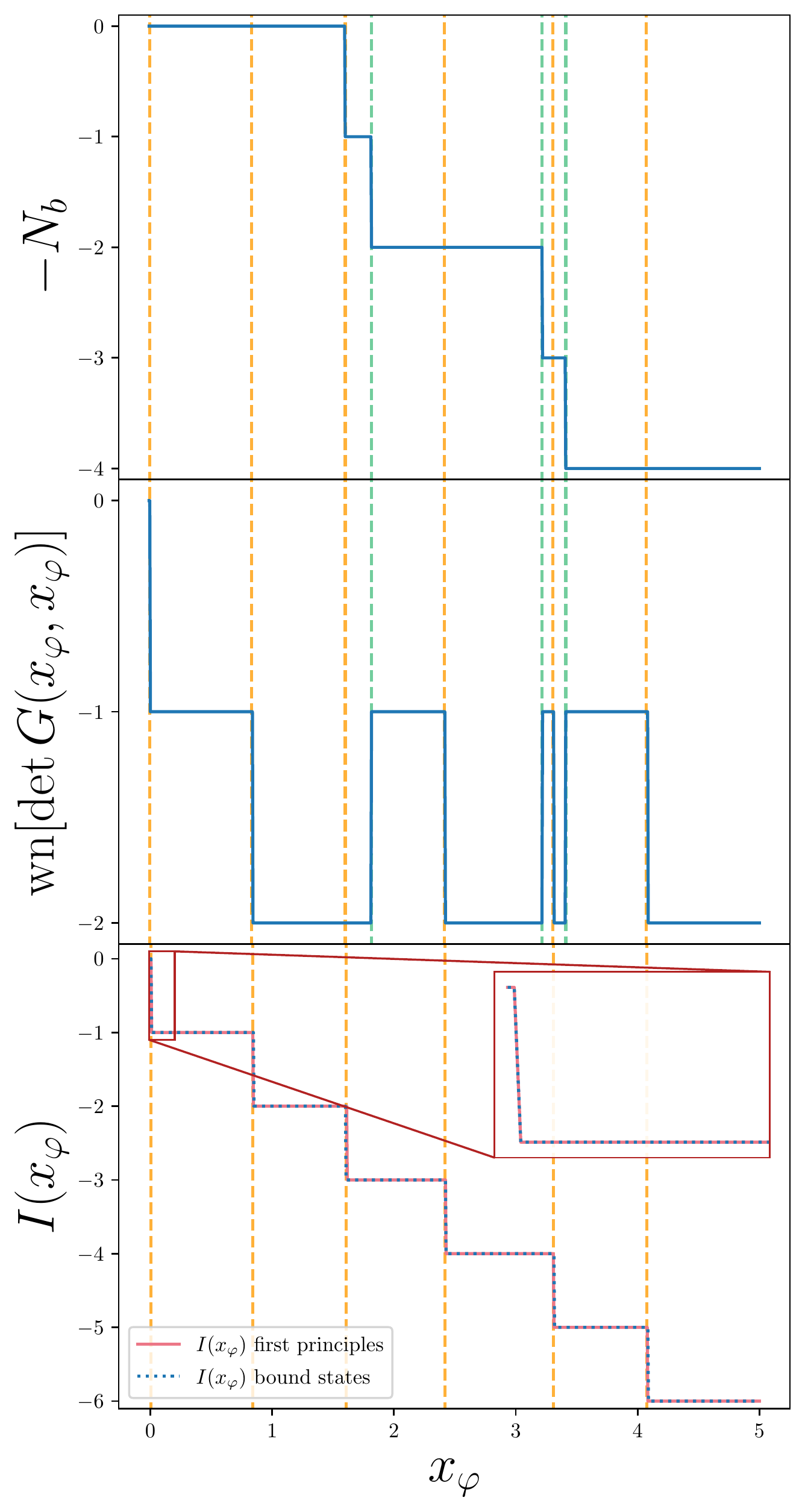}
                \caption{Top panel: negative of the number of bound states below the chemical potential $\mu=\frac{1}{2}(\max_{k}\epsilon_{6, k}+\min_{k}\epsilon_{7, k})$ inside a box of width $x_{\varphi}$. Middle panel: the winding number of boundary Green's function $G(x_{\varphi}, x_{\varphi})$. Bottom panel: comparison between the boundary invariants calculated from the first principles (\ref{I_repr1_0}) (in pink) and the one calculated with the help of the bound state representation (\ref{I_repr2_1}) (blue dotted line).}
                \label{fig: I_bs}
\end{figure}
\par
Finally, in Fig. \ref{fig: I_bs}, we present the results for the bound state representation of BI introduced in Sec. \ref{sec: bound_states}. The topmost panel of Fig. \ref{fig: I_bs} shows the first contribution to BI given by the formula (\ref{I_repr2_1}), that is, the negative of the number of bound states $-N_{b}$ inside the cavity (box) of width $x_{\varphi}$, lying below the chemical potential $\mu=\frac{1}{2}(\max_{k}\epsilon_{6, k}+\min_{k}\epsilon_{7, k})$. Although this number may be found on the basis of the formula (\ref{eq: num_bound_states}), we have found it numerically beneficial to calculate it via numerical diagonalization of the Hamiltonian (\ref{eq: hamiltonian}) in the basis of standing waves:
\begin{align}
    \chi_{n}(x)=\sqrt{\frac{2}{x_{\varphi}}}\sin\Bigg(\frac{\pi n x}{x_{\varphi}}\Bigg), \quad x\in[0, \ x_{\varphi}], \quad n\in\mathbb{N}.
\end{align}
The second contribution to BI defined by (\ref{I_repr2_1}), i.e. the winding number of the boundary Green's function (\ref{Dyson_main}) at $x=x'=x_{\varphi}$, is shown in the middle panel of Fig. \ref{fig: I_bs}. Although $-N_{b}$ is a strictly decreasing function of $x_{\varphi}$ (as one would expect on the physical grounds $N_{b}\sim 2x_{\varphi}\sqrt{\mu}/\pi$), $\text{wn}[G(x_{\varphi}, x_{\varphi})]$ shows three upward jumps by unity. Noteworthy, these are precisely cancelled by the corresponding downward jumps of $-N_{b}$ (as indicated by the punctured green lines) leading to the correct functional dependence of the boundary invariant as is demonstrated in the bottom panel of Fig. \ref{fig: I_bs}.

\section{Numerical results: From completely degenerate to generic regime}
\label{sec: near_degen}
\begin{figure*}[t]                \includegraphics[width=2\columnwidth]{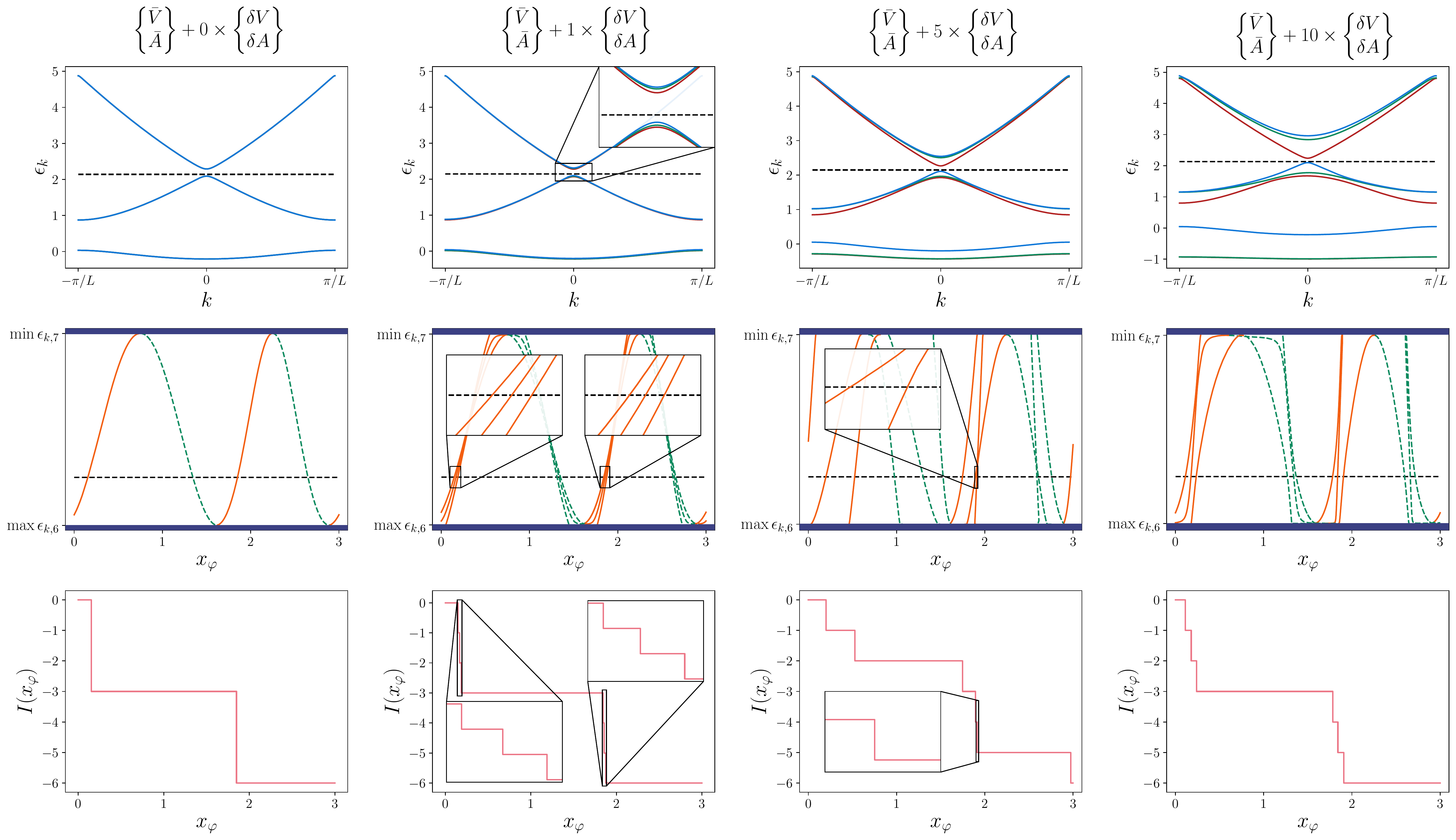}
                \caption{From top to bottom: energy spectrum, spectral flow of in-gap edge states, and boundary invariant for varying relative strengths of Abelian and non-Abelian parts of scalar and vector potentials. Here the strength of the perturbation increases from $0\times\begin{Bmatrix}\delta{V} \\ \delta{A}\end{Bmatrix}$ in the first column to $10\times\begin{Bmatrix}\delta{V} \\ \delta{A}\end{Bmatrix}$ in the last one. Here the chemical potential is indicated by the dashed black line and is chosen to lie a quarter of band-gap away from the top of the sixth energy band $\mu=\frac{1}{4}(3\max\epsilon_{k, 6}+\min\epsilon_{k, 6})$ (for any strength of the perturbation).}
                \label{fig: NC3}
\end{figure*}
In this section we are going to analyse the properties of $I$ and the associated spectral flow in the regimes ranging from the completely degenerate $\frac{||\delta{V}(x)||}{||\bar{V}(x)||}, \ \frac{||\delta{A}(x)||}{||\bar{A}(x)||}=0$, to the generic one $\frac{||\delta{V}(x)||}{||\bar{V}(x)||}, \ \frac{||\delta{A}(x)||}{||\bar{A}(x)||}\sim1$. Here $\bar{V}(x), \ \bar{A}(x)$ and $\delta{V}(x), \ \delta{A}(x)$ are the Abelian and non-Abelian parts of the corresponding potentials $V(x)=\bar{V}(x)+\delta{V}(x), \ A(x)=\bar{A}(x)+\delta{A}(x)$. In the following we consider the $N_{c}=3$ system defined by the potentials 
\begin{align}
    \bar{V}(x)&=(0.35\cos(2qx)+0.76\sin(qx))\lambda_{0},\\
    \bar{A}(x)&=(0.44\cos(3qx)+0.11\sin(2qx))\lambda_{0},\\
    \nonumber
    \delta{V}(x)&=0.074\lambda_{2}\cos(qx)+0.0108\lambda_{4}\cos(2qx)\\
    &+0.0043\lambda_{6}\cos(3qx)+0.0022\lambda_{8}\sin(qx),\\
    \nonumber
    \delta{A}(x)&=0.0031\lambda_{1}\sin(2qx)+0.011\lambda_{3}\sin(3qx)\\
    &+0.0065\lambda_{5}\sin(4qx)+0.0092\lambda_{7}\cos(4qx).
\end{align}
Here $\lambda_{0}$ is the $3\times 3$ identity matrix, $\lambda_{j}, \ j=1, ..., 8$ are the standard Gell-Mann matrices, and $q=\frac{2\pi }{L}$. In the following we assume that $L=3\ \text{a.u.}$ and as before we adopt the Hartree's atomic units. In the following we also assume that the chemical potential lies a quarter band-gap away from the top of the sixth energy band $\mu=\frac{1}{4}(3\max\epsilon_{k, 6}+\min\epsilon_{k, 6})$ independently of the perturbation strength. 
\par
Energy band structure, the spectral flow of edge states inside the gap, and the boundary invariant for the potentials $V(x)=\bar{V}(x)+0\times\delta{V}(x),\ \bar{V}(x)+1\times\delta{V}(x),\ \bar{V}(x)+5\times\delta{V}(x),\ \bar{V}(x)+10\times\delta{V}(x)$ (analogous for $A(x)$) are shown in Figure \ref{fig: NC3}. As the perturbation is turned on from zero, the energy sub-bands of a completely degenerate model slightly split apart, and as the strength of the perturbation increases further the sub-bands become more and more distinguishable. 
\par
Similarly, the edge state poles lying on top of one another in a model with degenerate bands are split apart by the perturbation. Let us note that as soon as the perturbation is turned on, the edge state dispersion's $x_{\varphi}$-gradient is no longer zero at the band edges (as it is \textit{always} the case in single-channel models [\onlinecite{Hatsugai_1993}, \onlinecite{Miles_21}]), instead the bound state poles enter the band at a slant. Touching the band edge by the edge state dispersion in the single channel case is an important property for establishing the conventional bulk-boundary correspondence in the spirit of Hatsugai [\onlinecite{Hatsugai_1993}]. It allows one to assign a certain sign-valued vorticity to the touching point, the sum of all vorticties over both band edges, bottom and top, representing the Chern index of the band. In the multichannel case this property does not hold, as discussed above. But it is nevertheless possible to establish the bulk-boundary correspondence by observing that the overall drop of the boundary invariant over the single $x_{\varphi}$-cycle coincides (up to the sign) with the number of the occupied bands $\nu$.
\par
As one can see, in a degenerate system, the boundary invariant features a pair of jumps by $-3$ corresponding to the escape of the particles of different "color charge" sitting in the same bound state. As the perturbation is switched on and the degeneracy is gone, the single step of $-3$ degenerates into a threesome of steps splitting further and further apart as the strength of the non-Abelian part of the potentials is increased.

\section{Low-energy theory}
\label{sec: LET}
In this section, we consider the class of systems with both Abelian and non-Abelian parts of vector and scalar potential being much less than the Fermi energy. In this case, it is possible to develop the so-called low-energy theory (LET) allowing one to get analytical insights into the physics behind the states residing near the chemical potential. 
\par
As per common practice [\onlinecite{Thakurathi_2018}, \onlinecite{Pletyukhov_2020_prr}, \onlinecite{Piasotski_2021}], present LET is developed on the basis of degenerate (Brillouin-Wigner) perturbation theory, where the periodic potentials generating the Fermi-surface resonances of $2k_{F, \bar\nu}=\frac{2\pi \bar\nu}{L}, \ \bar\nu=1, 2, ...$ are treated as the perturbation to the free particle Hamiltonian. In the following case, for simplicity purposes, we restrain ourselves to the systems satisfying
\begin{align}
 \int_{0}^{L}dxe^{-\frac{2\pi i\bar{\nu}x}{L}}V(x)\neq0, \quad   \int_{0}^{L}dxe^{-\frac{2\pi i\bar{\nu}x}{L}}A(x)\neq0.
\end{align}
That is we assume that the amplitudes of the mode corresponding to the Fermi surface $\bar\nu$ are non-zero and thus the gap is opened in the first order in perturbation theory. In the opposite scenario one simply has to consider higher-order perturbation theory to obtain the effective low-energy Hamiltonian (see Ref. [\onlinecite{Thakurathi_2018}, \onlinecite{Pletyukhov_2020_prr}] for details). 
\par
Note that in LET, the channel splitting of the energy bands is typically much smaller than "global" gaps generated purely by the periodicity of potentials, which are themselves much smaller than the Fermi energy. For this purpose, it is convenient to adopt the new labeling of states. In this section, the global gaps (from now on referred to as gaps for simplicity) are labeled by the index $\bar\nu=1, 2, ...$. In particular, when the chemical potential is placed in the gap $\bar\nu$, that means that $\nu=N_{c}\cdot\bar\nu$ bands are filled.

\subsection{Approximation for bulk Green's functions}
Consider either odd gap $\bar\nu = 2 p+1$ or even gap $\bar\nu=2 p$. In the LET description, an odd gap occurs when the two parabolas corresponding to the quadratic kinetic term in \eqref{Bloch_ham} and labelled by $l=p$ and $l=-p-1$ intersect each other at $k= \frac{\pi}{L}$, while an even gap occurs when the two parabolas labelled by $l=p$ and $l=-p$ intersect each other at $k=0$. To lift these degeneracies, it suffices to project the perturbative terms $\propto A_{l,l''}$ and $V_{l,l''}$ in \eqref{Bloch_ham} onto the quoted pairs of states. Thereby we get an effective low-energy Hamiltonian written in the $2 \times 2$ block form, with each block being a $N_c \times N_c$ matrix,
\begin{align}
    \mathcal{P} \left(h_k - \frac{k_F^2}{2 m} \right) \mathcal{P} &= h^{\text{eff}}_{\bar{k}} = \frac{\bar{k}^2}{2 m} +\frac{\bar{k}}{m} A_0 + V_0\nonumber \\
    & + \left(
    \begin{array}{cc}
         v_F \bar{k} + v_F A_0 &  \frac{\bar{k}}{m} A_{\bar\nu} + V_{\bar\nu} \\
         \frac{\bar{k}}{m} A_{\bar\nu}^{\dagger}  + V_{\bar\nu}^{\dagger} & -v_F \bar{k}  - v_F A_0 
    \end{array}\right).
    \label{LET_ham}
\end{align}
Hereby $\mathcal{P}$ denotes the corresponding projector, with the upper state being $p$, and the lower state being $-p-1$ ($-p$) for odd (even) $\bar\nu$.  In the following we label these states by $a,b=1,2$. We set the chemical potential $\mu= \frac{\pi^2 \bar\nu^2}{2 m L^2}$ and introduce the notations $\bar{k} = k - \frac{\pi}{L}$ ($\bar{k}=k$) for the odd (even) case as well as define the Fermi quasimomentum $k_F \equiv k_{F,\bar\nu} = \frac{\bar\nu \pi}{L}$ and  the Fermi velocity $v_F \equiv v_{F,\bar\nu} = \frac{\bar\nu \pi}{mL}$. In the above representation \eqref{LET_ham} we use the hermiticity property of $V(x)$ and $A (x)$ translated to their Fourier components \eqref{V_Fourier} and \eqref{A_Fourier}: $V_{-\bar\nu} = V_{\bar\nu}^{\dagger}$, $A_{-\bar\nu} = A_{\bar\nu}^{\dagger}$. Without loss of generality we may choose the zero-mode matrices $V_0$ and $A_0$ to be traceless.

Introducing $\bar{z} = z - \mu$ and neglecting the term $\frac{\bar{k}^2}{2m}$ we evaluate the corresponding effective Green's function $\bar{G}^{(0) \, \text{eff}}_{\bar{k}} = (\bar{z} -h^{\text{eff}}_{\bar{k}})^{-1}$.
The corresponding low-energy approximation for the Green's function in the spatial representation reads
 \begin{align}
     & G^{(0)} (x,x') \approx  \int_{-\infty}^{\infty}  \frac{\bar{d k} }{2 \pi} e^{i \bar{k} (x -x')} \nonumber \\
     & \times \left[ \bar{G}^{(0), \text{eff}}_{\bar{k},11} e^{i k_F (x -x')} +\bar{G}^{(0), \text{eff}}_{\bar{k},22} e^{-i k_F (x -x')} \right. \label{RWA} \\
     & \left. +\bar{G}^{(0), \text{eff}}_{\bar{k},12} e^{i k_F (x+x')}+\bar{G}^{(0), \text{eff}}_{\bar{k},21} e^{-i k_F (x+x')}\right]. \label{nonRWA}
 \end{align}

For the special single-channel case with $A (x) =0$ and $V_0=0$
\begin{align}
    \bar{G}^{(0), \text{eff}}_{\bar{k}} &= f (\bar{k},\bar{z}) \left( \begin{array}{cc}
         \bar{z} + v_F \bar{k} &  V_{\bar\nu} \\
         V_{\bar\nu}^* & \bar{z} - v_F \bar{k}, 
    \end{array}\right), \label{G0eff_sch} \\
    f (\bar{k},\bar{z}) &= \frac{1}{\bar{z}^2 - (v_F \bar{k})^2 - |V_{\bar\nu}|^2},
\end{align}
and
\begin{align}
     G^{(0)} (x,x) = 2 F (\bar{z}) [\bar{z} + |V_{\bar\nu}| \cos (2 k_F x + \varphi_{\bar\nu})],
\end{align}
where $\varphi_{\bar\nu}$ is the phase of the gap parameter $V_{\bar\nu}=|V_{\bar\nu}|e^{i\varphi_{\bar\nu}}$, $F (\bar{z}) = \int_{-\infty}^{\infty} \frac{d \bar{k}}{2 \pi} f (\bar{k},\bar{z}) = - \frac{1}{2 v_F} \frac{1}{\sqrt{|V_{\bar\nu}|^2-\bar{z}^2}}$, and in particular
\begin{align}
    F (\bar{z}= \omega + i \eta) = - \frac{1}{2 v_F} \begin{cases} \frac{i}{\sqrt{\omega^2 - |V_{\bar\nu}|^2}}, & \omega > |V_{\bar\nu}| \\
    \frac{1}{\sqrt{|V_{\bar\nu}|^2-\omega^2}}, & -|V_{\bar\nu}| < \omega < |V_{\bar\nu}|, \\
     - \frac{i}{\sqrt{\omega^2 - |V_{\bar\nu}|^2}}, & \omega <  -|V_{\bar\nu}|.
    \end{cases}
\end{align}
 
 \subsection{Boundary charge}
 \label{sec:QB_LET}

In the low-energy approximation we neglect the polarization charge $Q_P$, since it is $O (\frac{||V_{\bar\nu}||}{v_F k_F}\ln\frac{||V_{\bar\nu}||}{v_F k_F})$ (see Appendix \ref{scaling_polarization}). Our goal is to evaluate the boundary charge by the formula \eqref{QB_diff} making the LET approximation of $G^{(0)}$. Let us then consider
\begin{align}
   &\int_{x_{\varphi}}^{\infty} d x e^{-0^+ x} \text{tr} \{G^{(0)} (x,x_{\varphi}) [G^{(0)} (x_{\varphi},x_{\varphi})]^{-1} G^{(0)} (x_{\varphi},x)\} \nonumber \\
   &=  \text{tr} \Big\{ [G^{(0)} (x_{\varphi},x_{\varphi})]^{-1} \int \frac{d\bar{k}}{2 \pi}\int \frac{d\bar{k}'}{2 \pi}  \nonumber \\
   & \times \int_{x_{\varphi}}^{\infty} d x e^{i (\bar{k}-\bar{k}'+i0^+) (x-x_{\varphi})}  \nonumber \\
   \label{eq: strong_oscillating}
   & \times  \sum_{a,b,a'b'}  \bar{G}_{\bar{k}',b'a'}^{(0), \text{eff}} e^{i k_F (b' x_{\varphi} -a'x)} \bar{G}_{\bar{k},ab}^{(0), \text{eff}} e^{i k_F (a x-b x_{\varphi})}  \Big\}.
\end{align}
In order to suppress terms which are fast oscillating in $x$, we select only the terms in these sums, which have $a=a'$ (as it is  shown in Appendix \ref{sec: Strongly oscillating terms}, the rapidly oscillating terms give a $O (\frac{||V_{\bar\nu}||}{v_F k_F})$ contribution). (Here we use $a,b=1,2=+,-$). After that we perform the $x$-integration and obtain
\begin{align}
   &\int_{x_{\varphi}}^{\infty} dx e^{-0^+ x} \text{tr} \{G^{(0)} (x,x_{\varphi}) [G^{(0)} (x_{\varphi},x_{\varphi})]^{-1} G^{(0)} (x_{\varphi},x)\} \nonumber \\
    &= \text{tr} \Big\{  [G^{(0)} (x_{\varphi},x_{\varphi})]^{-1} \int \frac{d\bar{k}}{2 \pi}\int \frac{d\bar{k}'}{2 \pi}  \frac{i}{\bar{k}-\bar{k}'+i0^+} \nonumber \\
   & \times  \sum_{b,b'} \left[ \bar{G}_{\bar{k}'}^{(0),\ \text{eff}} \bar{G}_{\bar{k}}^{(0), \text{eff}}   \right]_{b'b} e^{i k_F (b'  -b )x_{\varphi}} \Big\}.
   \label{fried_low_energy}
\end{align}
Note the left-subsystem analog
\begin{align}
&\int_{-\infty} ^{x_{\varphi}} d x e^{0^+ x} \text{tr} \{G^{(0)} (x,x_{\varphi}) [G^{(0)} (x_{\varphi},x_{\varphi})]^{-1} G^{(0)} (x_{\varphi},x)\} \nonumber \\
    &= -\text{tr} \Big\{  [G^{(0)} (x_{\varphi},x_{\varphi})]^{-1} \int \frac{d\bar{k}}{2 \pi}\int \frac{d\bar{k}'}{2 \pi}  \frac{i}{\bar{k}-\bar{k}'-i0^+} \nonumber \\
   & \times  \sum_{b,b'} \left[ \bar{G}_{\bar{k}'}^{(0), \text{eff}} \bar{G}_{\bar{k}}^{(0), \text{eff}}   \right]_{b'b} e^{i k_F (b'  -b )x_{\varphi}} \Big\}.
   \label{fried_low_energy_left}
\end{align}

Representing 
\begin{align}
    &\bar{G}_{\bar{k}'}^{(0), \text{eff}} \bar{G}_{\bar{k}}^{(0), \text{eff}}  = \bar{G}_{\bar{k}'}^{(0), \text{eff}} \bar{G}_{\bar{k}'}^{(0), \text{eff}}  \nonumber \\
    &+\bar{G}_{\bar{k}'}^{(0), \text{eff}} (\bar{G}_{\bar{k}}^{(0), \text{eff}} -\bar{G}_{\bar{k}'}^{(0), \text{eff}})   \\
    &= - \frac{\partial}{\partial \bar{z}}  [\bar{G}_{\bar{k}'}^{(0), \text{eff}} ]
    \nonumber \\
    & + v_F (\bar{k}'-\bar{k}) \frac{\partial \bar{G}_{\bar{k}'}^{(0), \text{eff}}}{\partial \bar{z}} \sigma_z \bar{G}_{\bar{k}}^{(0), \text{eff}} 
\end{align}
on one hand, and
\begin{align}
    &\bar{G}_{\bar{k}'}^{(0), \text{eff}} \bar{G}_{\bar{k}}^{(0), \text{eff}}  = \bar{G}_{\bar{k}}^{(0), \text{eff}} \bar{G}_{\bar{k}}^{(0), \text{eff}}  \nonumber \\
    &+(\bar{G}_{\bar{k}'}^{(0), \text{eff}} -\bar{G}_{\bar{k}}^{(0), \text{eff}}) 
    \bar{G}_{\bar{k}}^{(0), \text{eff}}   \\
    &= - \frac{\partial}{\partial \bar{z}}  [\bar{G}_{\bar{k}}^{(0), \text{eff}} ]
    \nonumber \\
    &+ v_F (\bar{k}-\bar{k}' ) \bar{G}_{\bar{k}'}^{(0), \text{eff}} \sigma_z \frac{\partial \bar{G}_{\bar{k}}^{(0), \text{eff}}}{\partial \bar{z}}
\end{align}
on the other hand, we insert the symmetrized combination of the above expressions into \eqref{fried_low_energy}. Thereby we obtain
\begin{align}
     &\int_{x_{\varphi}}^{\infty} d x e^{-0^+x} \text{tr} \{G^{(0)} (x,x_{\varphi}) [G^{(0)} (x_{\varphi},x_{\varphi})]^{-1} G^{(0)} (x_{\varphi},x) \}\nonumber \\
     & \approx -\frac12 \text{tr} \left\{ [G^{(0)} (x_{\varphi},x_{\varphi})]^{-1}  \frac{\partial G^{(0)} (x_{\varphi},x_{\varphi})}{\partial \bar{z}} \right\} \\
   &-\frac{i}{2} v_F \text{tr} \Big\{ [G^{(0)} (x_{\varphi},x_{\varphi})]^{-1}  \int \frac{d\bar{k}}{2 \pi}\int \frac{d\bar{k}'}{2 \pi}  \nonumber \\
   & \times  \sum_{b,b'} \left[\frac{\partial \bar{G}_{\bar{k}'}^{(0), \text{eff}}}{\partial \bar{z}}  \sigma_z \bar{G}_{\bar{k}}^{(0), \text{eff}} - \bar{G}_{\bar{k}'}^{(0), \text{eff}} \sigma_z \frac{\partial \bar{G}_{\bar{k}}^{(0), \text{eff}}}{\partial \bar{z}} \right]_{b'b} \nonumber \\
   & \times e^{i k_F (b' x_{\varphi} -b x_{\varphi})} \Big\}.
   \label{non_wind}
\end{align}

\subsubsection{Single-channel case}

Noticing that
\begin{align}
    \bar{G}_{\bar{k}'}^{(0), \text{eff}}  \sigma_z \bar{G}_{\bar{k}}^{(0), \text{eff}} = \bar{G}_{\bar{k}}^{(0), \text{eff}}  \sigma_z \bar{G}_{\bar{k}'}^{(0), \text{eff}},
\end{align}
we show that
\begin{align}
\nonumber
&\frac{\partial \bar{G}_{\bar{k}'}^{(0), \text{eff}}}{\partial \bar{z}}  \sigma_z \bar{G}_{\bar{k}}^{(0), \text{eff}} - \bar{G}_{\bar{k}'}^{(0), \text{eff}} \sigma_z \frac{\partial \bar{G}_{\bar{k}}^{(0), \text{eff}}}{\partial \bar{z}} \\
&=  - 2 \bar{z} [f(\bar{k}',\bar{z}) - f (\bar{k},\bar{z})] \bar{G}_{\bar{k}'}^{(0), \text{eff}} \sigma_z \bar{G}_{\bar{k}}^{(0), \text{eff}} \label{int_term1} \\
&+f (\bar{k}',\bar{z}) \sigma_z \bar{G}_{\bar{k}}^{(0), \text{eff}}  - \bar{G}_{\bar{k}}^{(0), \text{eff}}\sigma_z f (\bar{k},\bar{z}). \label{int_term2}
\end{align}

Integrating over $\bar{k}$ and $\bar{k}'$, we observe that the term \eqref{int_term1} vanishes, while the term \eqref{int_term2} gives the contribution
\begin{align}
 2 [F (\bar{z})]^2 \left( \begin{array}{cc}
    0  &  V_{\bar\nu} \\
    -V_{\bar\nu}^*  & 0
 \end{array}\right).
\end{align}
Inserting this into \eqref{non_wind}, we obtain
\begin{align}
     &\int_{x_{\varphi}}^{\infty} d x e^{-0^+ x} G^{(0)} (x,x_{\varphi}) [G^{(0)} (x_{\varphi},x_{\varphi})]^{-1} G^{(0)} (x_{\varphi},x) \nonumber \\
     & \approx -\frac12 \frac{1}{\bar{z}+ |V_{\bar\nu}| \cos \bar{\varphi}_{\bar\nu}} - \frac12 [F (\bar{z})]^{-1} \frac{\partial F (\bar{z})}{\partial \bar{z}} \\
   &+ \frac{ v_F F (\bar{z}) |V_{\bar\nu}| \sin \bar{\varphi}_{\bar\nu}}{ \bar{z} + |V_{\bar\nu}| \cos \bar{\varphi}_{\bar\nu}} ,
   \label{non_wind1}
\end{align}
where $\bar{\varphi}_{\bar\nu} = 2 k_F x_{\varphi} + \varphi_{\bar\nu}$.

Integrating over the valence band states up to $-|V_{\bar\nu}|$, and accounting the pole at $\omega = - |V_{\bar\nu}| \cos \bar{\varphi}_{\bar\nu}$ in the band gap (setting $\mu = |V_{\bar\nu}|$), we obtain
\begin{align}
    Q_B (\bar{\varphi}_{\bar\nu}) &=  Q_B^{\text{corr}} -\frac14 + \frac12 \rcircleleftint \frac{d \bar{z}}{2 \pi i } \frac{1-2 v_F F (\bar{z}) |V_{\bar\nu}| \sin \bar{\varphi}_{\bar\nu}}{\bar{z} + |V_{\bar\nu}| \cos \bar{\varphi}_{\bar\nu}}  \nonumber \\
    &=  Q_B^{\text{corr}} - \frac14 +\Theta (\bar{\varphi}_{\bar\nu}) \nonumber \\
    &+ \frac{1}{2 \pi} \int_{-\infty}^{-|V_{\bar\nu}|} d \omega \frac{  |V_{\bar\nu}| \sin \bar{\varphi}_{\bar\nu}}{ (\omega + |V_{\bar\nu}| \cos \bar{\varphi}_{\bar\nu}) \sqrt{\omega^2 - |V_{\bar\nu}|^2}} .
\end{align}

Hereby the contribution $-\frac14$ arises from the half-winding of the function $F (z)$ around the branching point $\omega= -|V_{\nu}|$:
\begin{align}
\nonumber
    & \frac12 \rcircleleftint \frac{d \bar{z}}{2 \pi i }[F (\bar{z})]^{-1} \frac{\partial F (\bar{z})}{\partial \bar{z}} = \frac12 \rcircleleftint \frac{d \bar{z}}{2 \pi i } \frac{\bar{z}}{|V_{\bar\nu}|^2 -\bar{z}^2}  \\
    &= -\frac{1}{8 \pi i} \rcircleleftint_{-|V_{\bar\nu}|-i \eta}^{-|V_{\bar\nu}|+ i \eta} \frac{d \bar{z}}{|V_{\bar\nu}| +\bar{z}} -\text{Im} \int_{-\infty}^{-|V_{\bar\nu}|} \frac{d \omega}{2\pi } \frac{\omega + i \eta}{|V_{\bar\nu}|^2 - (\omega+i \eta)^2} \nonumber \\
    &= - \frac18 - \frac{1}{4 } \int_{-\infty}^{-|V_{\bar\nu}|} d \omega
    [\delta (\omega - |V_{\bar\nu}|) + \delta (\omega + |V_{\bar\nu}|)]=- \frac14.
\end{align}

In turn, the term $Q_B^{\text{corr}}$ accounts the contribution of deep lying occupied states which is missing after the spectrum linearization. Following the prescription of Ref.~[\onlinecite{Pletyukhov_2020_prr}],  we represent $Q_B^{\text{corr}}$ as a boundary charge in the gapless model with the nonlinearized spectrum ($\frac{\bar{k}^2}{2m}$ in the present case). The corresponding calculation, carried out in  Appendix \ref{app:Qcorr}, 
yields the result $Q_B^{\text{corr}}=-\frac14$.

Recalling that for $\varphi \in [-\pi,\pi]$ it holds
\begin{align}
     \int_{-\infty}^{-1} d x \frac{\sin \varphi}{(x + \cos \varphi) \sqrt{x^2-1}} = \varphi + \pi - 2 \pi \Theta (\varphi),
\end{align}
we finally obtain
\begin{align}
    Q_B (\bar{\varphi}_{\bar\nu}) = \frac{\bar{\varphi}_{\bar\nu}}{2 \pi},
\end{align}
which is consistent with our previous result for lattice models [\onlinecite{Pletyukhov_2020_prr}].

\subsubsection{Multi-channel case}
\label{sec: MC_LET}

Assuming that $||V_{\bar{\nu}}|| \sim v_F ||A_{\bar{\nu}}||$, we see that the scalar potential gives the dominant contribution to the off-diagonal component of the matrix in \eqref{LET_ham} at $\bar{k} \ll k_F$. Along with the additional simplifying assumption $A_0=0$, this results in the full neglect of the vector potential $A (x)$. Thereby we obtain the multichannel analog of \eqref{G0eff_sch}:
\begin{align}
    \bar{G}^{(0),\text{eff}}_{\bar{k}} &=  \left( \begin{array}{cc}
         f_1 (\bar{k},\bar{z}) (\bar{z} + v_F \bar{k}) &   f_1 (\bar{k},\bar{z}) V_{\bar\nu} \\
         V_{\bar\nu}^{\dagger} f_1 (\bar{k},\bar{z}) & (\bar{z} - v_F \bar{k}) f_2 (\bar{k},\bar{z}), 
    \end{array}\right),
\end{align}
where
\begin{align}
    f_1 (\bar{k},\bar{z}) &= \frac{1}{\bar{z}^2 - (v_F \bar{k})^2 - V_{\bar\nu} V_{\bar\nu}^{\dagger}}, \\
    f_2 (\bar{k},\bar{z}) &= \frac{1}{\bar{z}^2 - (v_F \bar{k})^2 - V_{\bar\nu}^{\dagger} V_{\bar\nu}}.
\end{align}
These matrices satisfy $f_1 V_{\bar\nu} = V_{\bar\nu} f_2$, $ V_{\bar\nu}^{\dagger} f_1= f_2 V_{\bar\nu}^{\dagger}$. Employing the singular-value decomposition $V_{\bar\nu} = U_1 \hat{V}_{\bar\nu} U_2^{\dagger}$ and $V_{\bar\nu}^{\dagger} = U_2 \hat{V}_{\bar\nu}^{*} U_1^{\dagger}$, where $U_{1,2}$ are unitary matrices, and $\hat{V}_{\bar\nu}$ is a diagonal matrix, we represent $f_1 = U_1 \hat{f} U_1^{\dagger}$ and $f_2 = U_2 \hat{f} U_2^{\dagger}$. Hereby
\begin{align}
    \hat{f} (\bar{k},\bar{z}) &= \frac{1}{\bar{z}^2 - (v_F \bar{k})^2 - |\hat{V}_{\bar\nu} |^2} 
\end{align}
is the diagonal matrix. Then
\begin{align}
    \bar{G}^{(0), \text{eff}}_{\bar{k}} &=  \left( \begin{array}{cc}
         U_1 \hat{f} U_1^{\dagger} (\bar{z} + v_F \bar{k}) &    U_1 \hat{f}  \hat{V}_{\bar\nu} U_2^{\dagger} \\
         U_2 \hat{V}_{\bar\nu}^{*} \hat{f} U_1^{\dagger}  & (\bar{z} - v_F \bar{k}) U_2 \hat{f} U_2^{\dagger}  
    \end{array}\right).
\end{align}
It follows
\begin{align}
    G^{(0)} (x,x) &= U_1 \bar{z} \hat{F} U_1^{\dagger} + U_2 \bar{z} \hat{F} U_2^{\dagger} \nonumber \\
    &+ U_1 \hat{F} \hat{V}_{\bar\nu} U_2^{\dagger} e^{2 i k_F x} + U_2  \hat{V}_{\bar\nu}^* \hat{F} U_1^{\dagger} e^{-2 i k_F x},
\end{align}
with the diagonal matrix $\hat{F} (\bar{z}) = \frac{1}{2 \pi} \int_{-\infty}^{\infty} d \bar{k} \hat{f} (\bar{k},\bar{z})$. In the off-diagonal representation we have
\begin{align}
    G^{(0)} (x,x) &= \bar{z} (F_1 + F_2) \nonumber \\
    &+ F_1 V_{\bar\nu} e^{2 i k_F x} + V_{\bar\nu}^{\dagger} F_1 e^{-2 i k_F x},
\end{align}
with $F_{1,2} (\bar{z}) = \frac{1}{2 \pi} \int_{-\infty}^{\infty} d \bar{k} f_{1,2} (\bar{k},\bar{z})$.

Denoting
\begin{align}
    \bar{G}_{av}^{(0)} &= \frac{1}{2 \pi} \int_{-\infty}^{\infty} d \bar{k} \bar{G}^{(0), \text{eff}}_{\bar{k}} = \left( \begin{array}{cc}
         U_1 \bar{z} \hat{F} U_1^{\dagger}  &    U_1 \hat{F}  \hat{V}_{\bar\nu} U_2^{\dagger} \\
         U_2 \hat{V}_{\bar\nu}^{*} \hat{F} U_1^{\dagger}  &  U_2 \bar{z} \hat{F} U_2^{\dagger} 
    \end{array}\right) \nonumber \\
    &=W \left( \begin{array}{cc}
         \bar{z} \hat{F}   &    \hat{F}  \hat{V}_{\bar\nu}  \\
          \hat{V}_{\bar\nu}^{*} \hat{F}  &  \bar{z} \hat{F} 
    \end{array}\right) W^{\dagger} \equiv W \hat{G}_{av}^{(0)} W^{\dagger},
\end{align}
where $W = \begin{pmatrix} U_1 & 0 \\ 0 & U_2 \end{pmatrix}$, we evaluate
\begin{align}
 &\int \frac{d\bar{k}}{2 \pi}\int \frac{d\bar{k}'}{2 \pi} \left[\frac{\partial \bar{G}_{\bar{k}'}^{(0), \text{eff}}}{\partial \bar{z}}  \sigma_z \bar{G}_{\bar{k}}^{(0), \text{eff}} - \bar{G}_{\bar{k}'}^{(0), \text{eff}} \sigma_z \frac{\partial \bar{G}_{\bar{k}}^{(0), \text{eff}}}{\partial \bar{z}} \right] \nonumber \\
 \nonumber
 &=W \left[ \frac{\partial \hat{G}_{av}^{(0)}}{\partial \bar{z}}  \sigma_z \hat{G}_{av}^{(0)} - \hat{G}_{av}^{(0)} \sigma_z \frac{\partial \hat{G}_{av}^{(0)}}{\partial \bar{z}} \right] W^{\dagger}  \\
 \nonumber
 &= W \left( \begin{array}{cc}
         0  &    2 \hat{F}^2  \hat{V}_{\bar\nu}  \\
          - 2 \hat{V}_{\bar\nu}^{*} \hat{F}^2  &  0
    \end{array}\right)W^{\dagger} \\
    &= \left( \begin{array}{cc}
         0  &    2 U_1 \hat{F}^2  \hat{V}_{\bar\nu}  U_2^{\dagger} \\
          - 2 U_2 \hat{V}_{\bar\nu}^{*} \hat{F}^2  U_1^{\dagger} &  0
    \end{array}\right).
\end{align}
Then
\begin{align}
& \int \frac{d\bar{k}}{2 \pi}\int \frac{d\bar{k}'}{2 \pi} \sum_{b,b'} e^{i k_F (b' x_{\varphi} -b x_{\varphi})} \nonumber \\
   & \times   \left[\frac{\partial \bar{G}_{\bar{k}'}^{(0), \text{eff}}}{\partial \bar{z}}  \sigma_z \bar{G}_{\bar{k}}^{(0), \text{eff}} - \bar{G}_{\bar{k}'}^{(0), \text{eff}} \sigma_z \frac{\partial \bar{G}_{\bar{k}}^{(0), \text{eff}}}{\partial \bar{z}} \right]_{b'b} \nonumber \\
   \nonumber
   & = 2 U_1 \hat{F}^2  \hat{V}_{\bar\nu}  U_2^{\dagger} e^{2 i k_F x_{\varphi}} -  2 U_2 \hat{V}_{\bar\nu}^{*} \hat{F}^2  U_1^{\dagger} e^{-2 i k_F x_{\varphi}} \\
   &= 2 F_1^2  V_{\bar\nu}  e^{2 i k_F x_{\varphi}} -  2 V_{\bar\nu}^{\dagger} F_1^2  e^{-2 i k_F x_{\varphi}}.
\end{align}

Introducing the following quantity
\begin{align}
    P (x_{\varphi})=& -i v_F \text{tr} \{ [G^{(0)} (x_{\varphi},x_{\varphi})]^{-1} \nonumber \\
    & \times ( F_1^2  V_{\bar\nu}  e^{2 i k_F x_{\varphi}}-   V_{\bar\nu}^{\dagger} F_1^2  e^{-2 i k_F x_{\varphi}})\},
    \label{P_def}
\end{align}
we show in Appendix \ref{app:polar_repr} that
\begin{align}
   \frac{\partial P (x_{\varphi})}{\partial x_{\varphi}}= - \frac{k_F}{v_F} \frac{\partial }{\partial \bar{z}} \text{tr} \{ [G^{(0)} (x_{\varphi},x_{\varphi})]^{-1}\}.
   \label{polar_conj}
\end{align}
Then
\begin{align}
    & Q_B (x_{\varphi}) = Q_B^{\text{corr}} \nonumber \\
    &+\frac{1}{2 \pi i} \rcircleleftint d \bar{z} \left[ \frac12  \frac{\partial }{\partial \bar{z}} \ln \det  G^{(0)} (x_{\varphi},x_{\varphi}) - P (x_{\varphi}) \right] \nonumber \\
    &=Q_B^{\text{corr}} + \frac{1}{2 \pi i} \rcircleleftint d \bar{z} \left[ \frac12  \frac{\partial }{\partial \bar{z}} \ln \det  G^{(0)} (x_{\varphi},x_{\varphi}) \right. \nonumber \\
    & \left. - P (0) + \frac{k_F}{v_F} \frac{\partial }{\partial \bar{z}} \int_0^{x_{\varphi}} d x \text{tr} \{ [G^{(0)} (x,x)]^{-1}\} \right],
\end{align}
where the contour embraces both poles and branch cuts in the counterclockwise direction, and $Q_B^{\text{corr}} = - \frac{N_c}{4}$. Hence
\begin{align}
    & Q_B (x_{\varphi}) - Q_B (0) \nonumber \\
    &= \rcircleleftint \frac{d \bar{z}}{2 \pi i} \frac{\partial }{\partial \bar{z}} \left[ \frac12   \ln \det  G^{(0)} (x_{\varphi},x_{\varphi})  - \frac12  \ln \det  G^{(0)} (0,0)  \right. \nonumber \\
    & \qquad \left.  + \frac{k_F}{v_F}  \int_0^{x_{\varphi}} d x \text{tr} \{ [G^{(0)} (x,x)]^{-1}\} \right]
    \label{QB_LET1} \\
    &= \frac{1}{2 \pi i} \rcircleleftint d \bar{z} q_B (x_{\varphi}; \bar{z}) - \frac{1}{2 \pi i} \rcircleleftint d \bar{z} q_B (0; \bar{z}).
    \label{QB_LET2}
\end{align}
In the representation \eqref{QB_LET2} we employed the boundary charge spectral density
\begin{align}
    q_B (x_{\varphi}; \bar{z}) = \frac12  \frac{\partial }{\partial \bar{z}} \ln \det  G^{(0)} (x_{\varphi},x_{\varphi}) - P (x_{\varphi}),
\end{align}
and \eqref{QB_LET2} holds true by virtue of the identity \eqref{polar_conj}.

The frequency contour integral gives the two contributions to $Q_B (x_{\varphi})-Q_B (0)$. Firstly, there is a branch cut along the real axis from $-\infty$ to $-\Delta_{\bar\nu} \equiv -\min_{1 \leq \lambda \leq N_c} |\hat{V}_{\bar\nu,\lambda}|$, that is up to the valence band edge determined by the smallest eigenvalue of $V_{\bar\nu}$. Secondly, there are poles lying in the bandgap below the chemical potential, that is in the range $-\Delta_{\bar\nu}< \bar{z} <0$. We should however keep in mind that these poles may correspond to bound states in either right- or left-subsystem.

To evaluate the branch-cut contribution, we use the representation \eqref{QB_LET1}, in which the integrand is expressed in terms of the full frequency derivative. Noticing that the phase of $\frac{\det G^{(0)} (x_{\varphi}, x_{\varphi})}{\det G^{(0)} (0,0)}$ is continuous at very large negative $\bar{z} \ll -\max_{1 \leq \lambda \leq N_c} |\hat{V}_{\bar\nu,\lambda}|$, while the jump in
\begin{align}
    [G^{(0)} (x,x)]^{-1} |_{\bar{z}=-\infty+i \eta} -[G^{(0)} (x,x)]^{-1} |_{\bar{z}=-\infty-i \eta} = 2 i v_F
\end{align}
contributes to
\begin{align}
    Q_B^{(b-c)} (x_{\varphi}) - Q_B^{(b-c)} (0) &= \frac{1}{2\pi i} \frac{k_F}{v_F}  \int_0^{x_{\varphi}} d x \text{tr} \{ 2 i v_F \}\\
    &= \frac{k_F}{\pi} N_c x_{\varphi} = \bar\nu N_c \frac{x_{\varphi}}{L}=\bar\rho x_{\varphi}.
\end{align}

In turn, to evaluate the poles' contribution to $Q_B (x_{\varphi})-Q_B (0)$, we use the representation \eqref{QB_LET2}. The poles $\bar{z}_j^{(p)}$ of $q_B (x_{\varphi; \bar{z}})$ are identified with the roots of the equation $\det G^{(0)} (x_{\varphi}, x_{\varphi}) =0 $, and thus we get
\begin{align}
    \frac{1}{2 \pi i} \rcircleleftint_{\text{poles}} d \bar{z} q_B (x_{\varphi}; \bar{z}) = \sum_j R_j (x_{\varphi}),
\end{align}
where we sum residua 
\begin{align}
\label{eq: resiidue}
    R_j (x_{\varphi}) = \lim_{\bar{z} \to \bar{z}_j^{(p)}} \frac12 \left( 1- \frac{ 2 P (x_{\varphi}) \det G^{(0)} (x_{\varphi},x_{\varphi}) }{\frac{\partial \det G^{(0)} (x_{\varphi},x_{\varphi})}{\partial \bar{z}}}\right)
\end{align}
of all the poles. It turns however out that $R_j (x_{\varphi}) = +1$ for the bound state in the right subsystem, and $R_j (x_{\varphi}) = 0$ for the bound state in the left subsystem (see Appendix \ref{app:res_pol} for the proof). Thus, we find the poles' contribution
\begin{align}
    Q_B^{(p)} (x_{\varphi}) - Q_B^{(p)} (0) = N_r (x_{\varphi}) - N_r (0)
\end{align}
in terms of the number $N_r (x_{\varphi})$ of the bound states in the right subsystem with energies lying in the gap below the chemical potential. In particular, in the single-channel case
\begin{align}
     N_r (x_{\varphi}) = \frac12 \left( 1 + \frac{\sin \bar{\varphi}_{\nu}}{| \sin \bar{\varphi}_{\bar\nu} |}\right) = \Theta (\sin \bar{\varphi}_{\bar\nu}),
\end{align}
which is in agreement with our previous result [\onlinecite{Pletyukhov_2020_prr}].

Finally,
\begin{align}
    Q_B (x_{\varphi}) - Q_B (0) = \bar\rho x_{\varphi} + N_r (x_{\varphi}) - N_r (0).
\end{align}
This formula allows for an immediate interpretation in terms of the spectral flow.
\par
Note that $x_{\varphi}$ can be re-defined $x_{\varphi} \to x_{\varphi} + \frac{1}{2 k_F N_c} \sum_{\lambda=1}^{N_c} \varphi_{\bar\nu, \lambda}$ to include the mean phase of the gap matrix eigenvalues.

\subsection{Comparison with exact theory}
\begin{figure}[t]                \includegraphics[width=\columnwidth]{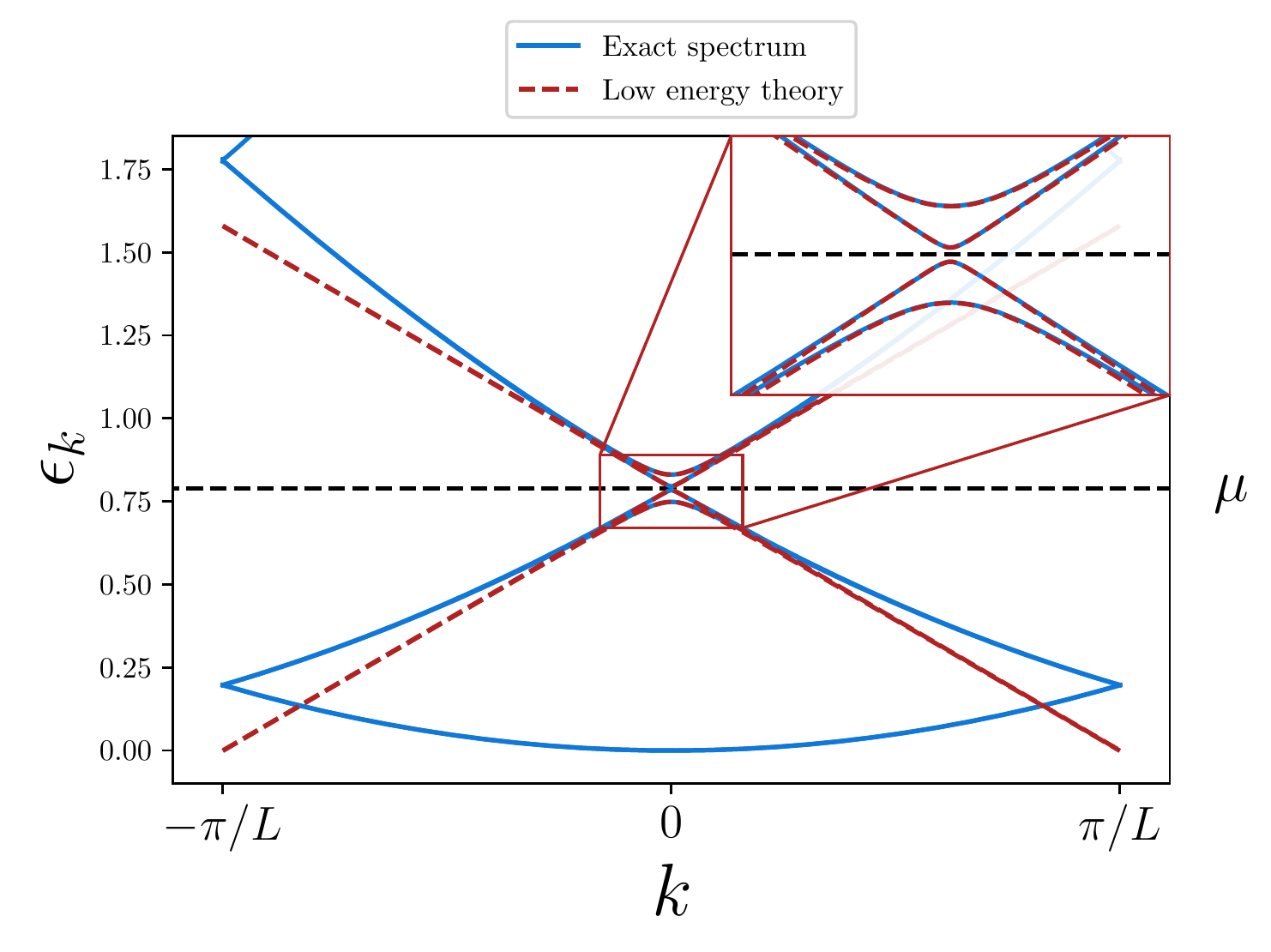}
                \caption{Comparison of the exact spectrum of the model defined by the potential (\ref{eq: pot_let}) (here $L=3 \ \text{a.u.}$) with its low-energy counterpart. As it is indicated in the inset the exact spectrum is well approximated by the low-energy one in the vicinity of the chemical potential located in the middle of the second energy gap $\mu=\frac{1}{2}(\max\epsilon_{k, 4}+\min\epsilon_{k, 5})$.}
                \label{fig: spectrum_LET}
\end{figure}
\begin{figure}[t]                \includegraphics[width=0.965\columnwidth]{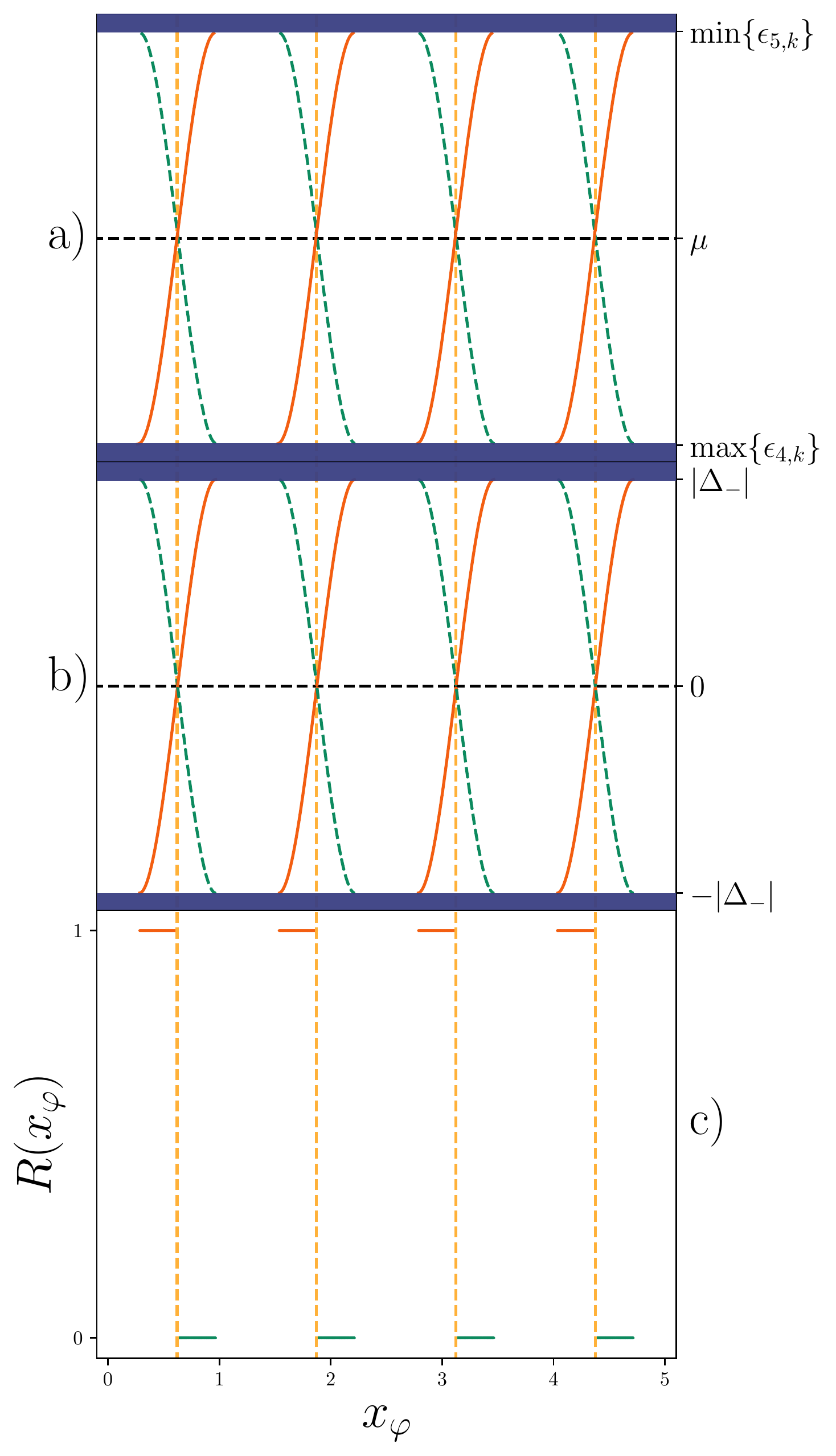}
                \caption{Comparison of the exact invariant with the one obtained on the basis of the low-energy theory. Panel a) shows the spectral flow of the in-gap edge states obtained within the exact theory. Panel b) shows the spectral flow of the in-gap edge states obtained within the low-energy approximation.
                Panel c) shows the $x_{\varphi}$ dependence of the residue $R(x_{\varphi})$ at the pole $\bar{z}_{+}^{*}$ (see \ref{eq: pole_LET}) lying below the chemical potential.}
                \label{fig: LET_vs_Exact}
\end{figure}
One expects the low-energy theory (LET) to be valid in the regime of small enough energy gaps $\frac{\Delta_{\bar\nu}}{\epsilon_{F}}\ll1$, where $\Delta_{\bar\nu}\sim\max_{1\leq\lambda\leq N_{c}}|\hat{V}_{\bar\nu, \lambda}|$ is the energy gap hosting the chemical potential, and $\epsilon_{F}$ is the Fermi energy. Thus, for the purpose of comparison of LET and exact result we have to choose a model with small enough potential amplitudes. In particular let us consider the following $N_{c}=2$ potential:
\begin{align}
\nonumber
    V(x)&=v_{0}\sigma_{0}\cos(3qx)+v_{1}\sigma_{1}\sin(2qx)\\
\label{eq: pot_let}
    &+v_{2}\sigma_{2}\sin(4qx)+v_{3}\sigma_{3}\cos(2qx).
\end{align}
Here as before $q=2\pi/L$, $\sigma_{0}$ and $\sigma_{j}$ are the identity and Pauli matrices respectively, and the potential amplitudes are chosen to be: $v_{0}=0.0065, \ v_{1}=0.035, \ v_{2}=0.00275, \ v_{3}=0.047$. In the following we additionally assume that $L=5  \ \text{a.u.}$ and the chemical potential is located in the middle of the second band gap $\mu=(\max\epsilon_{k, 4}+\min\epsilon_{k, 5})/2$ (that is $\bar\nu=2$). We can read off the effective potential:
\begin{align}
\label{eq: potential_LET}
    V_{2}=-i\alpha\sigma_{1}+\beta\sigma_{3}, \quad \alpha=\frac{v_{1}}{2}, \quad \beta=\frac{v_{3}}{2}.
\end{align}
The Dirac-type model features a quartet of energy bands with dispersion relations given by
\begin{align}
\label{eq: LET_disp}
\epsilon_{\bar{k}, \pm}^{(\sigma)}=\sigma\sqrt{v_{F}^{2}\bar{k}^{2}+\Delta_{\pm}^{2}}, \quad \Delta_{\pm}=\alpha\pm\beta.
\end{align}
Here $\sigma=-$ corresponds to the valence bands, $\sigma=+$ corresponds to the conduction ones, $\bar{k}=k-0=k$, and $v_{F}=k_{F}=\frac{\bar\nu\pi}{L}\equiv\frac{2\pi}{5}$. The comparison between the exact and approximate dispersion relations is shown in Fig. \ref{fig: spectrum_LET}. As one may infer from the inset, in the vicinity of the chemical potential, the two perfectly overlap.
\par
Performing the manipulations outlined in Section \ref{sec: MC_LET} we find that the LET approximation of the equal argument bare Green's function is given by
\begin{align}
\nonumber
    G^{(0)}(x, x)&=\bar{z}(\hat{F}_{+}+\hat{F}_{-})\sigma_{0}+(\Delta_{+}\hat{F}_{+}+\Delta_{-}\hat{F}_{-})\sin(2k_{F}x)\sigma_{1}\\
    &+(\Delta_{+}\hat{F}_{+}-\Delta_{-}\hat{F}_{-})\cos(2k_{F}x)\sigma_{3}.
\end{align}
Here $\hat{F}_{\pm}$ are the diagonal elements of $\hat{F}$ corresponding to the two different gaps $\Delta_{\pm}$. This immediately allows us to write the determinant of $G^{(0)}(x, x)$:
\begin{align}
\nonumber
    &\det G^{(0)}(x, x)\\
    &=2\hat{F}_{+}\hat{F}_{-}(\bar{z}^{2}+\Delta_{+}\Delta_{-}\cos(4k_{F}x))-\frac{1}{2v_{F}^{2}}.
\end{align}
It is straightforward to determine the poles inside the gap $\bar{z}\in[-|\Delta_{-}|, |\Delta_{-}|]$ satisfying $\det G^{(0)}(x, x)=0$. We find the following solutions
\begin{align}
\nonumber
\bar{z}_{\pm}^{*}&=\pm\frac{\Delta_{+}\Delta_{-}|\sin(4k_{F}x)|}{\sqrt{\Delta^{2}_{+}+\Delta_{-}^{2}+2\Delta_{+}\Delta_{-}\cos(4k_{F}x)}}\\
\label{eq: pole_LET}
&\times\Theta(|\Delta_{-}|/|\Delta_{+}|-\cos(4k_{F}x)),
\end{align}
where $\Theta$ is the Heaviside step function. The affiliation of a given pole (for a given value of $x$) with either left or right half-system follows from the sign of its spatial derivative: positive for the right half-system and negative for the left one. Using (\ref{P_def}) we obtain the following simple result for the $P$-function:
\begin{align}
\nonumber
    &\det G^{(0)}(x, x)P(x)\\
    &=2v_{F}\Delta_{+}\Delta_{-}\hat{F}_{+}\hat{F}_{-}(\hat{F}_{+}+\hat{F}_{-})\sin(4k_{F}x).
\end{align}
With this in hands, we readily find the following central result for the aforementioned model:
\begin{align}
\nonumber
    &Q_{B}^{(p)}(x_{\varphi})-Q_{B}^{(p)}(0)\\
    \nonumber
    &=\frac{\sin(4k_{F}x_{\varphi})}{|\sin(4k_{F}x_{\varphi})|}\Bigg(\frac{\Delta_{-}+\Delta_{+}\cos(4k_{F}{x}_{\varphi})\sin(4k_{F}x_{\varphi})}{|\Delta_{-}+\Delta_{+}\cos(4k_{F}{x}_{\varphi})\sin(4k_{F}x_{\varphi})|}\\
    \nonumber
    &\times\frac{\Delta_{+}(\Delta_{+}+\Delta_{-}\cos(4k_{F}x_{\varphi}))}{\sqrt{\Delta^{2}_{+}+\Delta_{-}^{2}+2\Delta_{+}\Delta_{-}\cos(4k_{F}x_{\varphi})}}-(\Delta_{-}\leftrightarrow\Delta_{+})\Bigg)\\
    \label{eq: LET_delta_QB}
    &\times\Theta(|\Delta_{-}|/|\Delta_{+}|-\cos(4k_{F}x_{\varphi})).
\end{align}
\par
The comparison of the results presented above with exact ones is shown in Fig. \ref{fig: LET_vs_Exact}. Panel a) shows the spectral flow of the in-gap edge states calculated from $\det[G^{(0)}(x_{\varphi}, x_{\varphi})]=0, \ \bar{z}\in[\max\epsilon_{k, 4}, \min\epsilon_{k, 5}]$, where $G^{(0)}(x_{\varphi}, x_{\varphi})$ is the exact Green's function for the model defined by the potential (\ref{eq: potential_LET}). On the other hand, the spectral flow based on the LET solution (\ref{eq: pole_LET}) is shown in panel b). As is indicated by dashed vertical yellow lines, the values of $x_{\varphi}$ where the edge states (belonging to the right half-system) escape the occupied part of the spectrum calculated on the basis of two approaches are in perfect agreement. Moreover, the entire lineshapes of the spectral flows are in exquisite tune with one another. This signifies that in the regime of small gaps, not only the exact spectrum in the vicinity of the chemical potential is well approximated by LET (see Fig. \ref{fig: spectrum_LET}), but also the Green's function itself. Let us also note that the discontinuity feature of the spectral flow (the poles of the left half-system do not continuously transform into the ones of the right half) discussed in Section \ref{sec: numerics_results} also persists in the low-energy regime. 
\par
Panel c) demonstrates the residue (\ref{eq: resiidue}) at the pole $\bar{z}^{*}_{+}$ lying below the chemical potential. As one can see, the residue takes on values $0$ or $1$ depending on the "chirality" of the pole. Whenever an edge state lies below the chemical potential, $R$ equals $0$ if the edge state is hosted by the left half system and equals $1$ if the edge state belongs to the right half system as it was pointed out in Sec. \ref{sec: MC_LET}. As is indicated by the dashed vertical yellow lines, the change in the value of the residue happens precisely at the points of the $x_{\varphi}$-cycle, where the right-half system's edge modes escape and the left-half system's enter the occupied part of the spectrum.

\section{Summary}
\label{sec: summary}
For generic continuum models with non-Abelian scalar and vector potentials, we have laid down a theoretical framework for boundary and interface charge investigations. Within our approach, both boundary and interface charges are expressed in terms of bulk position space Green's functions and its derivatives, allowing one to avoid the direct construction of the wavefunctions of an inhomogeneous system -- a hardly achievable goal in a system with composite energy bands (a typical scenario in multichannel systems).
\par
With the help of the Green's function representation of the boundary charge, we establish a topological integer-valued index $I(x_{\varphi})$ associated with the change in the boundary charge upon the translation of the entire system by $x_{\varphi}\in[0, L]$ towards the boundary. We demonstrated that $I(x_{\varphi})$ admits for two equivalent representations -- the winding number and the bound state representation. In the winding number representation, the boundary invariant is expressed via the winding number of a particular functional of bulk Green's functions as it encompasses the section of the real frequency axis corresponding to the occupied part of the energetic spectrum. On the other hand, in the bound state representation, the boundary invariant is expressed as the sum of the winding number of the boundary Green's function and the number of bound states hosted by the box of size $x_{\varphi}$ lying below the chemical potential.
\par
We observe that during a single $x_{\varphi}$-cycle, the boundary invariant exhibits exactly $\nu$ downward jumps by a unit of electron charge whenever the chemical potential is accommodated by the gap above the $\nu^{\text{th}}$ band. This naturally generalizes the results on the single-channel continuum models [\onlinecite{Miles_21}] to the multichannel realm. We find that the boundary invariant quantifies the spectral flow of the boundary eigenvalue problem: discontinuous jumps of $I(x_{\varphi})$ occur whenever an edge state pole escapes the occupied part of the spectrum -- that is, the unit of electron charge is carried away from the system in the process of spectral flow. Since the boundary invariant quantifies the boundary problem while being expressed in terms of bulk quantities solely, it presents itself as a symmetry-independent formulation of bulk-boundary correspondence in one-dimensional multichannel continuum models. Moreover, we have seen that the property of the bound state dispersion having zero $x_{\varphi}$-derivative at the band touching points [\onlinecite{Miles_21}] is not retained in the multichannel case. In the single-channel case this property is crucial for establishing the conventional bulk-boundary correspondence [\onlinecite{Hatsugai_1993}]. It allows one to assign an effective vorticity to the touching point, all vorticties over both band edges, bottom and top, summing up to the Chern index of the band. Even though in the multichannel case, this property does not hold, it is still possible to establish the bulk-boundary correspondence by observing that the net drop of $I$ over the single $x_{\varphi}$-cycle, up to the sign, coincides with the number of the occupied bands $\nu$ in the system. 
\par
Additionally, the quantization of excess charge accumulated on an isolated $\delta$-function (repulsive) impurity, the so-called interface charge $Q_{I}$, was demonstrated. In particular, we prove that the charge on the impurity is given by the winding number and thus, is a topological invariant. Via this construction, we provide a rigorous mathematical proof of the nearsightedness principle in the general one-dimensional multichannel continuum models (see Ref. [\onlinecite{Muller_2021}] for an analogous result for tight-binding lattice models). We observe that the charge piled up on the impurity is restricted to the values $-N_{c}, ..., -1, 0$, generalizing the results of Ref. [\onlinecite{Miles_21}]. 
\par
For the systems with potential amplitudes being much smaller than the typical value of kinetic energy, we additionally develop a Green's function-based low-energy theory (LET) of boundary and interface charges. In particular, for the models with scalar potential only, we provide a rigorous proof that the change in $Q_{B}$ upon a continuous shift of a lattice towards the boundary by $x_{\varphi}$ is, up to integer contributions, a perfectly linear function of $x_{\varphi}$. This development provides the multichannel generalization of the field-theoretical version of the surface charge theorem discovered in Ref. [\onlinecite{Piasotski_2021}] for single-channel tight-binding models. We reveal that the agreement between LET and exact results becomes exquisite in the regime of small energy gaps (potential amplitudes), signifying that in this regime, in the vicinity of the chemical potential, the exact Green's function of the theory is perfectly approximated by its LET counterpart. Of future interest is the generalization of the existing low-energy theory to the one-dimensional systems with non-Abelian vector potentials.

\section{Acknowledgments}
We acknowledge durable exchange of ideas on the subject of boundary and interface charges with J. Klinovaja and D. Loss. KP is grateful to A. Samson and S. Miles for enlightening discussions on numerical methods. This work was supported by the Deutsche Forschungsgemeinschaft via RTG 1995. Simulations were performed with computing resources granted by RWTH Aachen University.

\begin{appendix}

\section{Derivation of (\ref{ident_int_G0G0}). (\ref{GG_half}), and (\ref{ident_int_GG})}
\label{app:int_G0G0}

Performing the sum over unit cell index we obtain
\begin{align}
  & \int_{-\infty}^{\infty} d x \,\, G^{(0)} (0,x) G^{(0)} (x,0) \nonumber \\
=&  \sum_{n=-\infty}^{\infty}  \left( \frac{L}{2 \pi} \right)^2 \int_{-\pi/L}^{\pi/L} d k' \, \int_{-\pi/L}^{\pi/L} d k \, e^{i (k-k' ) n L} \nonumber \\
& \times \int_{0}^{ L} d \bar{x} \,\,  G_{k'}^{(0)} (0,\bar{x}) G_k^{(0)} (\bar{x}, 0) \nonumber   \\
=& \frac{L}{2 \pi}   \int_{-\pi/L}^{\pi/L} d k \,  \int_{0}^{ L} d \bar{x} \,\,  G_{k}^{(0)} (0,\bar{x}) G_k^{(0)} (\bar{x}, 0) .
\end{align}
Exploiting the Fourier representation \eqref{fourier_repr} we further observe that the above expression equals
\begin{align}
& \frac{L}{2 \pi}   \int_{-\pi/L}^{\pi/L} d k \,  \frac{1}{L^2} \nonumber \\
& \times \sum_{l_1,l'_1,l_2,l'_2} \int_{0}^{ L} d \bar{x} \,\,  \bar{G}_{k,l_1 l'_1}^{(0)} e^{-\frac{2\pi i}{L} l'_1\bar{x} } e^{\frac{2\pi i}{L} l_2 \bar{x} } \bar{G}_{k,l_2 l'_2}^{(0)} \nonumber \\
=& \frac{L}{2 \pi}   \int_{-\pi/L}^{\pi/L} d k \,  \frac{1}{L} \sum_{l_1,l'_2} \,\,  \left( \bar{G}_{k}^{(0)}   \bar{G}_{k}^{(0)} \right)_{l_1,l'_2} \nonumber \\
=& - \frac{\partial}{\partial \omega} \left[  \frac{L}{2 \pi}   \int_{-\pi/L}^{\pi/L} d k \,  \frac{1}{L} \sum_{l_1,l'_2} \,\,  \left(   \bar{G}_{k}^{(0)} \right)_{l_1,l'_2} \right] \nonumber \\
=&  - \frac{\partial}{\partial \omega} G^{(0)} (0,0) .
\end{align}

We also prove \eqref{GG_half} which is a single-sided analog of \eqref{ident_int_G0G0}. Without loss of generality we set $x_0=0$, since the coordinate systems can be chosen arbitrarily. Summing up over the unit cells, we obtain
\begin{align}
    &\int_0^{\infty} d x e^{-0^+ x} G^{(0)} (0,x) G^{(0)} (x,0) \nonumber \\
    &= \int_0^L d \bar{x} \left( \frac{L}{2 \pi}\right)^2 \int_{-\pi/L}^{\pi/L} d k'  \int_{-\pi/L}^{\pi/L} d k \nonumber  \\ & \times \sum_{n=0}^{\infty} e^{i (k-k'+i 0^+) n L} G_{k'}^{(0)} (0,\bar{x}) G_k^{(0)} (\bar{x},0) \nonumber \\
    &= \left( \frac{L}{2 \pi}\right)^2 \int_{-\pi/L}^{\pi/L} d k'  \int_{-\pi/L}^{\pi/L} d k \frac{1}{1-e^{i (k-k'+i 0^+) L}} \nonumber \\
    & \times  \int_0^L d \bar{x} G_{k'}^{(0)} (0,\bar{x}) G_k^{(0)} (\bar{x},0).
    \label{GG_k_repr}
\end{align}
To evaluate the latter integral, we differentiate \eqref{eq_G0k} with respect to $\omega$
\begin{align}
G_k^{(0)} (\bar{x},0)+ (z - H_{\bar{x}}^{\rightarrow})  \frac{\partial G_k^{(0)} (\bar{x},0)}{\partial \omega} = 0 . \label{eq_G0k_der_omega}
\end{align}
Multiplying the result with $ G_{k'}^{(0)} (0,\bar{x})$ and integrating over $\bar{x}$ from $0^+$ to $L^-$, we obtain with help of \eqref{eq_G0kH}
\begin{align}
     &  \int_0^L d \bar{x} G_{k'}^{(0)} (0,\bar{x}) G_k^{(0)} (\bar{x},0)
     \nonumber \\
     &= -\int_{0^+}^{L^-} d \bar{x} G_{k'}^{(0)} (0,\bar{x}) (H_{\bar{x}}^{\leftarrow}-H_{\bar{x}}^{\rightarrow}) \frac{\partial G_k^{(0)} (\bar{x},0)}{\partial \omega} \nonumber \\
     &= \frac{1}{2m} \Bigg[ G_{2,k'}^{(0)} (0,\bar{x}) \frac{\partial G_k^{(0)} (\bar{x},0)}{\partial \omega}  -  G_{k'}^{(0)} (0,\bar{x}) \frac{\partial G_{1,k}^{(0)} (\bar{x},0) }{\partial \omega}
     \nonumber \\
     & \qquad -2 i  G_{k'}^{(0)} (0,\bar{x}) A (\bar{x}) G_k^{(0)}  (\bar{x},0)\Bigg]_{0^+}^{L^-}.
\end{align}
Taking into account the boundary conditions \eqref{PBC1}, \eqref{PBC2} as well as the property \eqref{discont_der}, we derive
\begin{align}
     & 2 m  \int_0^L d \bar{x} G_{k'}^{(0)} (0,\bar{x}) G_k^{(0)} (\bar{x},0)
     \nonumber \\
     &=  \Bigg[ G_{2,k'}^{(0)} (0,0^+) \frac{\partial G_k^{(0)} (0,0)}{\partial \omega}  - G_{k'}^{(0)} (0,0) \frac{\partial G_{1,k}^{(0)} (0^+,0) }{\partial \omega}
     \nonumber \\
     & \qquad -2 i  G_{k'}^{(0)} (0,0) A (0) \frac{\partial G_k^{(0)}  (0,0)}{\partial \omega}\Bigg] (e^{i (k -k')L}-1) \nonumber \\
     &-2m e^{i (k-k') L} \frac{\partial G_k^{(0)} (0,0)}{\partial \omega}.
     \label{GG_partial}
\end{align}
Since
\begin{align}
& \frac{L}{2 \pi} \int_{-\pi/L}^{\pi/L} dk' \frac{e^{i (k-k') L}}{1-e^{i (k-k'+i 0^+)}} \nonumber \\
&= -1 + \frac{1}{2 \pi i}\oint_{k'\in[-\pi/L, \pi/L]}  \frac{d (e^{ik'L})}{e^{i k' L}-e^{i k L} e^{-0^+ L}} =0,
\end{align}
the last term of \eqref{GG_partial} does not give any contribution to \eqref{GG_k_repr}. The remaining terms of \eqref{GG_partial} yield
\begin{align}
    & -2 m  \int_0^{\infty} d x e^{-0^+ x} G^{(0)} (0,x) G^{(0)} (x,0) \nonumber \\
    &= \left( \frac{L}{2 \pi}\right)^2 \int_{-\pi/L}^{\pi/L} d k'  \int_{-\pi/L}^{\pi/L} d k \nonumber \\
    & \times \Bigg[ G_{2,k'}^{(0)} (0,0^+) \frac{\partial G_k^{(0)} (0,0)}{\partial \omega}  -  G_{k'}^{(0)} (0,0) \frac{\partial G_{1,k}^{(0)} (0^+,0) }{\partial \omega}
     \nonumber \\
     & \qquad -2 i  G_{k'}^{(0)} (0,0) A (0) \frac{\partial G_k^{(0)}  (0,0)}{\partial \omega}\Bigg],
\end{align}
which proves \eqref{GG_half} for $x_0 =0$.

Finally, to prove \eqref{ident_int_GG}, we notice that \eqref{ident_int_G0G0} can be easily generalized to
\begin{align}
  \int_{-\infty}^{\infty} d x \,\, G^{(0)} (x'',x) G^{(0)} (x,x') = - \frac{\partial G^{(0)} (x'',x')}{\partial \omega}.
\end{align}
Then, using the equation \eqref{Dyson_main} we show that
 \begin{align}
&   \int_{-\infty}^{\infty} d x \,  G (x_{\varphi},x) G (x,x_{\varphi})  = - \frac{\partial G^{(0)} (x_{\varphi}, x_{\varphi})}{\partial \omega} \nonumber \\
&+ G^{(0)} (x_{\varphi},0)  [G^{(0)} (0,0)]^{-1} \frac{\partial G^{(0)} (0, x_{\varphi})}{\partial \omega} \nonumber \\
&+ \frac{\partial G^{(0)} (x_{\varphi},0)}{\partial \omega} [G^{(0)} (0,0)]^{-1}  G^{(0)} (0,x_{\varphi}) \nonumber \\
&+ G^{(0)} (x_{\varphi},0)    \frac{\partial [G^{(0)} (0,0)]^{-1} }{\partial \omega  }  G^{(0)} (0,x_{\varphi}) \nonumber \\
&=  - \frac{\partial G (x_{\varphi}, x_{\varphi})}{\partial \omega}.
 \end{align}

\section{Proof of the identities (\ref{L_herm}) and (\ref{G12_ident})}
\label{app:main_ident}

We consider the Green's function
\begin{align}
    G_{x_0} (x,x') &= G^{(0)} (x,x') \nonumber \\
    &- G^{(0)} (x,x_0) [G^{(0)} (x_0,x_0)]^{-1} G^{(0)} (x_0,x'),
    \label{GF_bound_arb}
\end{align}
which corresponds to a model with the infinite-strength delta-impurity at $x=x_0$, and for $x_0=0$ it coincides with \eqref{GF_bound}. 

For the arguments' values $x<x_0$ and $x'>x_0$ this function identically vanishes, $G_{x_0} (x,x') \equiv 0$, since such points belong to the different subsystems, between which there is no communication. Differentiating the right-hand side of \eqref{GF_bound_arb} with respect to both $x$ and $x'$, and then taking the limits $x \to x_0^-$ and $x' \to x_0^+$, we obtain
\begin{align}
    G^{(0)} (x_0^-,x_0^+) = G_1^{(0)} (x_0^-,x_0) [G^{(0)} (x_0,x_0)]^{-1} G_2^{(0)} (x_0,x_0^+).
    \label{G12_mp}
\end{align}
This proves the identity \eqref{G12_ident}.

Analogously we establish that
\begin{align}
    G^{(0)} (x_0^+,x_0^-) = G_1^{(0)} (x_0^+,x_0) [G^{(0)} (x_0,x_0)]^{-1} G_2^{(0)} (x_0,x_0^-).
    \label{G12_pm}
\end{align}
Comparing the right-hand sides of \eqref{G12_mp} and \eqref{G12_pm} with each other, and accounting the properties \eqref{discont_der} and \eqref{cont_2nd_der}, we establish
 \begin{align}
      & G_1^{(0)} (x_0^{\pm},x_0)   [G^{(0)} (x_0,x_0)]^{-1} +i  A (x_0) \nonumber \\
      =& [G^{(0)} (x_0,x_0)]^{-1} G_2^{(0)} (x_0,x_0^{\pm}) -i  A (x_0).
\end{align}
For the upper sign choice, this relation coincides with \eqref{L_herm}.

\section{Derivation of (\ref{dQP})}
\label{app:Delta_QP}

To derive (\ref{dQP}), we consider
\begin{align}
 Q_P (x_{\varphi}) =& - \int_{x_{\varphi}}^{L+x_{\varphi}} d x \frac{x}{L} [\rho^{(0)} (x ) - \bar{\rho}] \nonumber \\
 =& Q_P (0)  - \int_{L}^{L+x_{\varphi}} d x \frac{x}{L} [\rho^{(0)} (x ) - \bar{\rho}] \nonumber \\
 & +  \int_0^{x_{\varphi}} d x \frac{x}{L} [\rho^{(0)} (x ) - \bar{\rho}] \nonumber \\
 =& Q_P (0)  - \int_{0}^{x_{\varphi}} d x \frac{x+L}{L} [\rho^{(0)} (x ) - \bar{\rho}] \nonumber \\
 & +  \int_0^{x_{\varphi}} d x \frac{x}{L} [\rho^{(0)} (x ) - \bar{\rho}] \nonumber \\
 =& Q_P (0)  - \int_{0}^{x_{\varphi}} d x  [\rho^{(0)} (x ) - \bar{\rho}] \nonumber \\
 =& Q_P (0)  + \bar{\rho} x_{\varphi} - \int_{0}^{x_{\varphi}} d x  \rho^{(0)} (x ) .
\end{align}

\section{Average density in the bulk}
\label{ap: av_des}
Here we would like to prove that whenever the chemical potential is located above the gap number $\nu$, the average density in the bulk amounts to $\bar\rho=\frac{\nu}{L}$. In order to proceed, we review some elementary properties of the bulk eigenstates of the system in close analogy with [\onlinecite{Miles_21}]. In particular, the eigenstates satisfy the Bloch condition
\begin{align}
    \psi_{k, \alpha}(x+L)=e^{ikL}\psi_{k, \alpha}(x), 
\end{align}
and are normalized to unity over the unit cell
\begin{align}
\nonumber
    \int_{0}^{L}dx\psi_{k, \alpha}^{\dagger}(x)\psi_{k, \alpha}(x)&=\sum_{\lambda=1}^{N_{c}}\int_{0}^{L}dx\psi_{k, \alpha}^{\lambda \dagger}(x)\psi_{k, \alpha}^{\lambda}(x)\\
    \label{eq: normaliz}
    &=1.
\end{align}
Furthermore, the following completeness and identity resolution identities hold
\begin{align}
\frac{L}{2\pi}\int_{-\pi/L}^{\pi/L}\psi_{k, \alpha}(x)\psi^{\dagger}_{k, \alpha}(x')&=1_{N_{c}}\delta(x-x'), \\
\frac{L}{2\pi}\int{dx}\psi_{k, \alpha}^{\dagger}(x)\psi_{k', \alpha'}(x)&=\delta(k-k')\delta_{\alpha, \alpha'}.
\end{align}
With this in hands we can represent the bulk single-particle Green's function as
\begin{align}
\label{eq: spec_repr}
G^{(0)}(x, x')=\frac{L}{2\pi}\sum_{\alpha}\int_{-\pi/L}^{\pi/L}dk\frac{\psi_{k, \alpha}(x)\psi^{\dagger}_{k, \alpha}(x')}{\omega-\epsilon_{k, \alpha}+i\eta}.
\end{align}
Using the representation \eqref{eq: spec_repr} as well as the normalization condition \eqref{eq: normaliz}, we get from the definition \eqref{eq: density_def_av}
\begin{align}
\nonumber
\bar\rho&=\frac{1}{L}\int_{0}^{L}dx\rho^{(0)}(x)\\
\nonumber
&=-\frac{1}{\pi}\frac{1}{L}\frac{L}{2\pi}\text{Im}\int{d\omega}\Theta(\mu-\omega)\int_{0}^{L}dx\text{tr}\{G^{(0)}(x, x)\}\\
\nonumber
&=\frac{1}{L}\frac{L}{2\pi}\sum_{\alpha}\int_{-\pi/L}^{\pi/L}dk\int{d\omega}\Theta(\mu-\omega)\\
\nonumber
&\times\int_{0}^{L}dx\text{tr}\{\psi_{k, \alpha}(x)\psi^{\dagger}_{k, \alpha}(x)\}\delta(\omega-\epsilon_{k, \alpha})\\
\nonumber
&=\frac{1}{L}\frac{L}{2\pi}\sum_{\alpha}\int_{-\pi/L}^{\pi/L}dk\int{d\omega}\Theta(\mu-\omega)\\
\nonumber
&\times\int_{0}^{L}dx\psi^{\dagger}_{k, \alpha}(x)\psi_{k, \alpha}(x)\delta(\omega-\epsilon_{k, \alpha})\\
&=\frac{1}{L}\sum_{\alpha}\int{d\omega}\Theta(\mu-\omega)\delta(\omega-\epsilon_{k, \alpha})=\frac{\nu}{L},
\end{align}
where $\nu$ is the number of occupied bands below the chemical potential.

\section{Scaling of $Q_{P}$ and strongly oscillating terms in low-energy theory}
\subsection{Polarization charge}
\label{scaling_polarization}
We start our discussion with the evaluation of the bulk density (\ref{eq: density_def}) in the low-energy limit
\begin{align}
\label{eq: rho_ap}
    \rho^{(0)}(x)=-\frac{1}{\pi}\text{Im}\int_{-\infty}^{0}d\bar{z}\text{tr}G^{(0)}(x, x).
\end{align}
In the low-energy approximation we get
\begin{align}
\nonumber
    &\text{Im}[\text{tr}G^{(0)}(x, x)]\\
    &=2\sum_{\lambda=1}^{N_{c}}(\bar{z}+\text{Re}[\tilde{U}_{\lambda\lambda}e^{i\varphi_{\bar\nu, \lambda}}e^{2ik_{F}x}]|\hat{V}_{\bar\nu, \lambda}|)\text{Im}[\hat{F}_{\lambda}(\bar{z})],
\end{align}
where we have defined $\tilde{U}=U_{2}^{\dagger}U_{1}$. In the following we are only interested in $\delta\rho^{(0)}(x)=\rho^{(0)}(x)-\bar\rho$. Averaging (\ref{eq: rho_ap}) over the unit cell and using $k_{F}=\frac{\bar\nu\pi}{L}$ we obtain 
\begin{align}
    \bar\rho=-\frac{2}{\pi}\sum_{\lambda=1}^{N_{c}}\int_{-\infty}^{0}d\bar{z}\bar{z}\text{Im}[\hat{F}_{\lambda}(\bar{z})] .
\end{align}
It thus follows
\begin{align}
\label{eq: rho_ap}
    \delta\rho^{(0)}(x)&=-\frac{2}{\pi}\sum_{\lambda=1}^{N_{c}}\text{Re}[\tilde{U}_{\lambda\lambda}e^{i\varphi_{\bar\nu, \lambda}}e^{2ik_{F}x}]|\hat{V}_{\bar\nu, \lambda}|\\
    &\times\int_{-\infty}^{0}d\bar{z}\text{Im}[\hat{F}_{\lambda}(\bar{z})].
\end{align}
As it is usually the case in $1+1$ dimensional field-theory, the integrals featuring only a single propagator have to be complemented with the high-energy cutoff $\Lambda_{\epsilon}\simeq\epsilon_{F}\simeq v_{F}k_{F}$. Bearing this in mind we write 
\begin{align}
\nonumber
&-\frac{1}{\pi}\int_{-\infty}^{0}d\bar{z}\text{Im}[\hat{F}_{\lambda}(\bar{z})]\simeq -\frac{1}{\pi}\int_{-\Lambda_{\epsilon}}^{0}d\bar{z}\text{Im}[\hat{F}_{\lambda}(\bar{z})]\\
&=-\frac{1}{2\pi v_{F}}\int_{-\frac{\Lambda_{\epsilon}}{|\hat{V}_{\bar\nu, \lambda}|}}^{-1}d\bar\omega\frac{1}{\sqrt{\bar\omega^{2}-1}}\simeq \frac{1}{2\pi v_{F}}\ln\frac{|\hat{V}_{\bar\nu, \lambda}|}{\Lambda_{\epsilon}}.
\end{align}
Hence we conclude that 
\begin{align}
    &\rho^{(0)}(x)\simeq\sum_{\lambda=1}^{N_{c}}(\text{Re}[\tilde{U}_{\lambda\lambda}e^{i\varphi_{\bar\nu, \lambda}}e^{2ik_{F}x}]|\hat{V}_{\bar\nu, \lambda}|)\frac{\ln\frac{|\hat{V}_{\bar\nu, \lambda}|}{\Lambda_{\epsilon}}}{\pi v_{F}}.
\end{align}
Substituting this result into the definition of the polarization charge (\ref{eq: polarization_def}) and performing the integral we find 
\begin{align}
\nonumber
    Q_{P}&=-\frac{1}{\pi}\sum_{\lambda=1}^{N_{c}}\text{Im}[\tilde{U}_{\lambda\lambda}e^{i\varphi_{\bar\nu, \lambda}}]\frac{|\hat{V}_{\bar\nu,  \lambda}|}{2v_{F}k_{F}}\ln\frac{|\hat{V}_{\bar\nu,  \lambda}|}{2\Lambda_{E}}\\
    &=O\Bigg(\frac{D}{v_{F}k_{F}}\ln\frac{D}{v_{F}k_{F}}\Bigg),
\end{align}
where $D=\text{max}_{1\leq \lambda \leq N_{c}}|\hat{V}_{\bar\nu,  \lambda}|$. This justifies the assertion made in Section \ref{sec:QB_LET}.

\subsection{Strongly oscillating terms in (\ref{eq: strong_oscillating})}
\label{sec: Strongly oscillating terms}
The fast oscillating contribution to (\ref{eq: strong_oscillating}) is given by
\begin{align}
   &FO= \frac{1}{\pi}\text{Im}\int_{-\infty}^{0}d\bar{z}\text{tr} \Big\{ [G^{(0)} (x_{\varphi},x_{\varphi})]^{-1} \int \frac{d\bar{k}}{2 \pi}\int \frac{d\bar{k}'}{2 \pi}  \nonumber \\
   & \times \int_{x_{\varphi}}^{\infty} d x e^{i (\bar{k}-\bar{k}'+i0^+) (x-x_{\varphi})}  \nonumber \\
   & \times \sum_{a,b,b'}  \bar{G}_{\bar{k}',b'a}^{(0), \text{eff}}\bar{G}_{\bar{k},\bar{a}b}^{(0), \text{eff}} e^{i k_F (b'-b) x_{\varphi}}e^{-2ik_{F}ax}\Big\} \nonumber\\
   &= \frac{1}{\pi}\text{Im}\int_{-\infty}^{0}d\bar{z}\text{tr} \Big\{ [G^{(0)} (x_{\varphi},x_{\varphi})]^{-1} \int \frac{d\bar{k}}{2 \pi}\int \frac{d\bar{k}'}{2 \pi}  \nonumber \\
   & \times i\sum_{a,b,b'} \frac{\bar{G}_{\bar{k}',b'a}^{(0), \text{eff}}\bar{G}_{\bar{k},\bar{a}b}^{(0), \text{eff}} e^{i k_F (b'-b-2a) x_{\varphi}}}{ \bar{k}-\bar{k}'-2ak_{F}+i\eta}\Big\}, \quad \bar{a}=-a.
\end{align}
To the leading order in $\frac{D}{v_{F}k_{F}}$ we have 
\begin{align}
   &FO \approx - \frac{1}{2 k_F \pi}\text{Im}\int_{-\Lambda_c}^{-D}d\bar{z}\text{tr} \Big\{ \frac{i \bar{z}}{F_1 (\bar{z}) + F_2 (\bar{z})} \nonumber \\
   & \times  (F_1^2 (\bar{z}) e^{-2 i k_F x_{\varphi}}  - F_2^2 (\bar{z}) e^{2 i k_F x_{\varphi}} )  \Big\} \nonumber \\
   & \approx - \frac{N_c  \sin (2 k_F x_{\varphi})}{4 v_F k_F \pi} \int_{-\Lambda_c}^{-D}d\bar{z}  \,  \frac{\bar{z}}{\sqrt{\bar{z}^2 -D^2}} \nonumber \\
   &= O \Bigg( \frac{D}{v_F k_F} \Bigg).
\end{align}

\section{High-energy correction to the low-energy approximation of the boundary charge}
\label{app:Qcorr}

Let us compute the correction term $Q_B^{\text{corr}}$ as the boundary charge in the gapless model with the parabolic spectrum. Its bulk Green's function  reads
\begin{align}
\nonumber
    G^{(p)} (x,x') &= \int_{-\infty}^{\infty} \frac{d k}{2 \pi} \frac{e^{i k (x-x')}}{\omega+ i \eta - \frac{k^2}{2m}} \\
    &= - i m  \frac{e^{i |x-x'| \sqrt{2 m (\omega+i \eta)}}}{ \sqrt{2m (\omega + i \eta)}}.
\end{align}
Inserting this into the general formula \eqref{QB_diff} we evaluate
\begin{align}
    Q_B^{\text{corr}} &=\frac{1}{\pi} \text{Im} \int_{-\infty}^{\mu} d \omega  \int_0^{\infty} d x \left\{- i m  \frac{e^{2 i x \sqrt{2 m (\omega+i \eta)}}}{ \sqrt{2m (\omega + i \eta)}} \right\} \nonumber  \\
    &= \frac{1}{4\pi} \text{Im} \int_{-\infty}^{\mu} \frac{d \omega}{\omega + i \eta} = - \frac14.
\end{align}

\section{Proof of (\ref{polar_conj})}
\label{app:polar_repr}

First, we notice the useful relations
\begin{align}
    F_1^2 = \frac{1}{4 v_F^2} \frac{1}{V_{\bar\nu} V_{\bar\nu}^{\dagger} -\bar{z}^2}, \quad F_2^2 = \frac{1}{4 v_F^2} \frac{1}{V_{\bar\nu}^{\dagger} V_{\bar\nu} -\bar{z}^2},
\end{align}
and
\begin{align}
     \frac{\partial F_1}{\partial \bar{z}} &= 4 v_F^2 \bar{z} F_1^3 , \quad  \frac{\partial (\bar{z} F_1)}{\partial \bar{z}}  =   4 v_F^2 V_{\bar\nu} V_{\bar\nu}^{\dagger}  F_1^3, \\
      \frac{\partial F_2}{\partial \bar{z}} &= 4 v_F^2 \bar{z} F_2^3 , \quad  \frac{\partial (\bar{z} F_2)}{\partial \bar{z}}  =   4 v_F^2 V_{\bar\nu}^{\dagger} V_{\bar\nu}  F_2^3.
\end{align}

On the basis of \eqref{P_def} we evaluate
\begin{align}
    & \frac{1}{2 v_F k_F} \frac{\partial P (x_{\varphi})}{\partial x_{\varphi}}\nonumber \\
    =& \text{tr} \{ [G^{(0)} (x_{\varphi},x_{\varphi})]^{-1} (  F_1^2  V_{\bar\nu}  e^{2 i k_F x_{\varphi}}+    V_{\bar\nu}^{\dagger} F_1^2  e^{-2 i k_F x_{\varphi}})\} \nonumber \\
    -& \text{tr} \{ [G^{(0)} (x_{\varphi},x_{\varphi})]^{-1} \nonumber \\
    & \times ( F_1  V_{\bar\nu}  e^{2 i k_F x_{\varphi}}-   V_{\bar\nu}^{\dagger} F_1  e^{-2 i k_F x_{\varphi}}) [G^{(0)} (x_{\varphi},x_{\varphi})]^{-1} \nonumber \\
    & \times ( F_1^2  V_{\bar\nu}  e^{2 i k_F x_{\varphi}}-   V_{\bar\nu}^{\dagger} F_1^2  e^{-2 i k_F x_{\varphi}})\},
\end{align}
where we employed the relation $\frac{\partial G^{-1}}{\partial x_{\varphi}} = - G^{-1} \frac{\partial G}{\partial x_{\varphi}} G^{-1}$. Transforming further, we obtain
\begin{align}
     &\frac{1}{2 v_F k_F} \frac{\partial P (x_{\varphi})}{\partial x_{\varphi}}\nonumber \\
     =&  \text{tr} \{  F_1^2  V_{\bar\nu}  e^{2 i k_F x_{\varphi}} [G^{(0)} (x_{\varphi},x_{\varphi})]^{-1} \nonumber \\
    & \times  (\bar{z} F_1 + \bar{z} F_2 + 2  V_{\bar\nu}^{\dagger} F_1  e^{-2 i k_F x_{\varphi}})  [G^{(0)} (x_{\varphi},x_{\varphi})]^{-1}  \} \nonumber \\
    +& \text{tr} \{   V_{\bar\nu}^{\dagger} F_1^2  e^{-2 i k_F x_{\varphi}} [G^{(0)} (x_{\varphi},x_{\varphi})]^{-1} \nonumber \\
    & \times (\bar{z} F_1 + \bar{z} F_2 + 2 F_1  V_{\bar\nu}  e^{2 i k_F x_{\varphi}})  [G^{(0)} (x_{\varphi},x_{\varphi})]^{-1}  \} .
\end{align}

Let us now consider
\begin{align}
    & \frac{1}{2 v_F k_F} \left[ \frac{\partial P (x_{\varphi})}{\partial x_{\varphi}}+ \frac{k_F}{v_F} \frac{\partial }{\partial \bar{z}} \text{tr} \{ [G^{(0)} (x_{\varphi},x_{\varphi})]^{-1}\} \right] \nonumber \\
    =&  \text{tr} \{  F_1^2  V_{\bar\nu}  e^{2 i k_F x_{\varphi}} [G^{(0)} (x_{\varphi},x_{\varphi})]^{-1} \nonumber \\
    & \times  (\bar{z} F_1 + \bar{z} F_2 + 2  V_{\bar\nu}^{\dagger} F_1  e^{-2 i k_F x_{\varphi}})  [G^{(0)} (x_{\varphi},x_{\varphi})]^{-1}  \} \nonumber \\
    +&  \text{tr} \{   V_{\bar\nu}^{\dagger} F_1^2  e^{-2 i k_F x_{\varphi}} [G^{(0)} (x_{\varphi},x_{\varphi})]^{-1} \nonumber \\
    & \times (\bar{z} F_1 + \bar{z} F_2 + 2 F_1  V_{\bar\nu}  e^{2 i k_F x_{\varphi}})  [G^{(0)} (x_{\varphi},x_{\varphi})]^{-1}  \} \nonumber \\
    -& 2  \text{tr} \{ [G^{(0)} (x_{\varphi},x_{\varphi})]^{-1} (V_{\bar\nu}V_{\bar\nu}^{\dagger}F_1^3 + F_2^3 V_{\bar\nu}^{\dagger} V_{\bar\nu}  \nonumber \\
    &  +  \bar{z} F_1^3 V_{\bar\nu} e^{2 i k_F x_{\varphi}} + V_{\bar\nu}^{\dagger} \bar{z} F_1^3 e^{-2 i k_F x_{\varphi}} ) [G^{(0)} (x_{\varphi},x_{\varphi})]^{-1} \} .
\end{align}
Shortening the notation $G^{(0)} (x_{\varphi},x_{\varphi})=G$ and re-arranging terms in the above expression, we find its alternative representation
\begin{align}
     =&  \text{tr} \{ [ (\bar{z}  F_1  + V_{\bar\nu}^{\dagger} F_1 e^{-2 i k_F x_{\varphi}})G -G (\bar{z}  F_2+ V_{\bar\nu}^{\dagger} F_1 e^{-2 i k_F x_{\varphi}} )] \nonumber \\
     &\times   e^{2 i k_F x_{\varphi}} G^{-1} (  G^{-1} F_1^2 V_{\bar\nu}-F_1^2 V_{\bar\nu} G^{-1} )G^{-1}\}  \nonumber \\
     +& \text{tr} \{ [ (\bar{z}  F_2  +F_1 V_{\bar\nu}  e^{2 i k_F x_{\varphi}})G - G (\bar{z} F_1  +F_1 V_{\bar\nu} e^{2 i k_F x_{\varphi}})] \nonumber \\
    &\times e^{-2 i k_F x_{\varphi}}  G^{-1} (G^{-1}  V_{\bar\nu}^{\dagger} F_1^2-V_{\bar\nu}^{\dagger} F_1^2 G^{-1})G^{-1}\}  .
    \label{tr_vanish}
\end{align}
By virtue of the identities
\begin{align}
    (\bar{z} F_1 + V_{\bar\nu}^{\dagger} F_1 e^{-2 i k_F x_{\varphi}}) G - G (z F_2 + V_{\bar\nu}^{\dagger} F_1 e^{-2 i k_F x_{\varphi}}) &=0  , \\
    (\bar{z} F_2 + F_1 V_{\bar\nu} e^{2 i k_F x_{\varphi}}) G- G (\bar{z} F_1 + F_1 V_{\bar\nu} e^{2 i k_F x_{\varphi}}) &= 0,
\end{align}
which can be checked by a direct calculation, the terms in \eqref{tr_vanish} are mutually cancelling, and thus \eqref{polar_conj} is proven.

\section{Residua values for the bound state in right and left subsystems}
\label{app:res_pol}

Suppose that a root $\bar{z}_r^{(p)}$ of the equation $\det G^{(0)} (x_{\varphi}, x_{\varphi}) =0$ gives a pole of the boundary Green's function $ G (x,x; \bar{z})$, which corresponds to a bound state in the right subsystem $x> x_{\varphi}$. In the vicinity of the pole  we approximate
\begin{align}
    & G (x,x; \bar{z}) \approx - \frac{1}{\bar{z} - \bar{z}_r^{(p)}} \frac{1}{\partial_{\bar{z}} \det G^{(0)} (x_{\varphi},x_{\varphi}; \bar{z}_r^{(p)})} \nonumber \\
    & \times G^{(0)} (x,x_{\varphi}; \bar{z}_r^{(p)}) \text{adj}[G^{(0)} (x_{\varphi},x_{\varphi}; \bar{z}_r^{(p)})] G^{(0)} (x_{\varphi},x; \bar{z}_r^{p}) \nonumber\\
    &= \frac{| \psi_{e,r} (x) \rangle \langle \psi_{e,r} (x) |}{\bar{z} - \bar{z}_r^{(p)}}.
    \label{BGF_pole_approx}
\end{align}
Hereby $\text{adj} [G^{(0)}] = \det G^{(0)} \, [G^{(0)}]^{-1}$ denotes the adjugate matrix, which is regular at the pole frequency.

The normalization condition of the edge state $| \psi_{e,r} (x) \rangle$ in the right subsystem
\begin{align}
    \int_{x_{\varphi}}^{\infty} d x \langle \psi_{e,r} (x) |  \psi_{e,r} (x) \rangle =1
\end{align}
implies that
\begin{align}
    1 =&  - \frac{1}{\partial_{\bar{z}} \det G^{(0)} (x_{\varphi},x_{\varphi}; \bar{z}_r^{(p)})} \text{tr} \left\{ \text{adj}[G^{(0)} (x_{\varphi},x_{\varphi}; \bar{z}_r^{(p)})] \right. \nonumber \\
   & \left. \times \int_{x_{\varphi}}^{\infty} d x e^{- 0^+ x}  G^{(0)} (x_{\varphi},x; \bar{z}_r^{p}) G^{(0)} (x, x_{\varphi}; \bar{z}_r^{p})\right\}.
\end{align}
Using the analogous integral evaluation in Section \ref{sec:QB_LET}, we obtain
\begin{align}
    1 =&  - \frac{1}{\partial_{\bar{z}} \det G^{(0)} (x_{\varphi},x_{\varphi}; \bar{z}_r^{(p)})}  \nonumber \\
   &  \times \Big( -\frac12 \partial_{\bar{z}}  \det G^{(0)} (x_{\varphi},x_{\varphi}; \bar{z}_r^{(p)})  \nonumber \\
   &  +P (x_{\varphi}; \bar{z}_r^{(p)}) \det G^{(0)} (x_{\varphi},x_{\varphi}; \bar{z}_r^{(p)}) \Big)  ,
\end{align}
which gives the residue value
\begin{align}
     R_r (x_{\varphi}) =&   \frac12  \left( 1 - \frac{2 P (x_{\varphi}; \bar{z}_r^{(p)}) \det G^{(0)} (x_{\varphi},x_{\varphi}; \bar{z}_r^{(p)}) }{\partial_{\bar{z}} \det G^{(0)} (x_{\varphi},x_{\varphi}; \bar{z}_r^{(p)})} \right) =1.
\end{align}

In turn, for a root $\bar{z}_l^{(p)}$ of the equation $\det G^{(0)} (x_{\varphi}, x_{\varphi}) =0$ corresponding to the edges state $| \psi_{e,l} (x) \rangle$ in the left subsystem $x < x_{\varphi}$, we have an approximation of the boundary Green's function  in the vicinity of $\bar{z} \approx \bar{z}_l^{(p)}$, which is similar to \eqref{BGF_pole_approx}. But the normalization condition
\begin{align}
    1 =&  \int_{-\infty}^{x_{\varphi}} d x \langle \psi_{e,l} (x) |  \psi_{e,l} (x) \rangle \nonumber \\
    =& - \frac{1}{\partial_{\bar{z}} \det G^{(0)} (x_{\varphi},x_{\varphi}; \bar{z}_l^{(p)})} \text{tr} \left\{ \text{adj}[G^{(0)} (x_{\varphi},x_{\varphi}; \bar{z}_l^{(p)})] \right. \nonumber \\
   & \left. \times \int_{-\infty}^{x_{\varphi}} d x e^{ 0^+ x}  G^{(0)} (x_{\varphi},x; \bar{z}_l^{p}) G^{(0)} (x, x_{\varphi}; \bar{z}_l^{p})\right\}
\end{align}
contains the complementary integration range $(-\infty, x_{\varphi}]$. Employing in this case \eqref{fried_low_energy_left} instead of \eqref{fried_low_energy}, we eventually obtain
\begin{align}
    1 &=- \frac{1}{\partial_{\bar{z}} \det G^{(0)} (x_{\varphi},x_{\varphi}; \bar{z}_l^{(p)})}  \nonumber \\
   &  \times \Big( -\frac12 \partial_{\bar{z}}  \det G^{(0)} (x_{\varphi},x_{\varphi}; \bar{z}_l^{(p)})  \nonumber \\
   &  - P (x_{\varphi}; \bar{z}_l^{(p)}) \det G^{(0)} (x_{\varphi},x_{\varphi}; \bar{z}_l^{(p)}) \Big)  ,
\end{align} 
which gives the residue value
\begin{align}
     R_l (x_{\varphi}) =&   \frac12  \left( 1 - \frac{2 P (x_{\varphi}; \bar{z}_l^{(p)}) \det G^{(0)} (x_{\varphi},x_{\varphi}; \bar{z}_l^{(p)}) }{\partial_{\bar{z}} \det G^{(0)} (x_{\varphi},x_{\varphi}; \bar{z}_l^{(p)})} \right) =0.
\end{align}

\end{appendix}

\end{document}